\documentclass[fleqn,usenatbib]{mnras}

\usepackage{newtxtext,newtxmath}

\usepackage[T1]{fontenc}
\usepackage{ae,aecompl}

\usepackage{graphicx}	
\usepackage{amsmath}	
\usepackage{bm}
\usepackage{dblfloatfix}
\usepackage{ulem}
\usepackage{multirow}

\usepackage{scalerel,tikz}
\usetikzlibrary{svg.path}
\definecolor{orcidlogocol}{HTML}{A6CE39}
\tikzset{orcidlogo/.pic={
 \fill[orcidlogocol] svg{M256,128c0,70.7-57.3,128-128,128C57.3,256,0,198.7,0,128C0,57.3,57.3,0,128,0C198.7,0,256,57.3,256,128z};
 \fill[white] svg{M86.3,186.2H70.9V79.1h15.4v48.4V186.2z}
 svg{M108.9,79.1h41.6c39.6,0,57,28.3,57,53.6c0,27.5-21.5,53.6-56.8,53.6h-41.8V79.1z M124.3,172.4h24.5c34.9,0,42.9-26.5,42.9-39.7c0-21.5-13.7-39.7-43.7-39.7h-23.7V172.4z}
 svg{M88.7,56.8c0,5.5-4.5,10.1-10.1,10.1c-5.6,0-10.1-4.6-10.1-10.1c0-5.6,4.5-10.1,10.1-10.1C84.2,46.7,88.7,51.3,88.7,56.8z};
}}
\newcommand\orcidicon[1]{\href{https://orcid.org/#1}{\mbox{\scalerel*{
\begin{tikzpicture}[yscale=-1,transform shape]
\pic{orcidlogo};
\end{tikzpicture}
}{|}}}}

\title[HMF]{The dark matter halo mass function in the $\Lambda$CDM cosmology at all times and over all scales -- from planetary to galaxy cluster masses}

\author[H. Zheng et al.]{%
Haonan Zheng\orcidicon{0000-0002-1665-5138}$^{1,2,3}$\thanks{Email: hnzheng@pku.edu.cn}, 
Sownak Bose\orcidicon{0000-0002-0974-5266}$^{2}$\thanks{Email: sownak.bose@durham.ac.uk}, 
Carlos S. Frenk\orcidicon{0000-0002-2338-716X}$^{2}$, 
Liang Gao\orcidicon{0000-0002-9276-917X}$^{4,5,6}$,  
Adrian Jenkins\orcidicon{0000-0003-4389-2232}$^{2}$,
\newauthor 
Shihong Liao\orcidicon{0000-0001-7075-6098}$^{3}$, 
Yizhou Liu\orcidicon{0009-0005-8855-0748}$^{4,5}$,
Volker Springel\orcidicon{0000-0001-5976-4599}$^{7}$, 
Jie Wang$^{3,4,8}$, 
Simon D. M. White\orcidicon{0000-0002-1061-6154}$^{7}$ 
\vspace*{0.1cm}\\%
$^{1}$Kavli Institute for Astronomy and Astrophysics, Peking University, Beijing 100871, China\\%
$^{2}$Institute for Computational Cosmology, Department of Physics, University of Durham, South Road, Durham, DH1 3LE, UK\\%
$^{3}$Key Laboratory for Computational Astrophysics, National Astronomical Observatories, Chinese Academy of Sciences, Beijing 100101, China\\%
$^{4}$Institute for Frontiers in Astronomy and Astrophysics, Beijing Normal University, Beijing 102206, China\\%
$^{5}$School of Physics and Astronomy, Beijing Normal University, Beijing 100875, China\\%
$^{6}$School of Physics and Laboratory of Zhongyuan Light, Zhengzhou University, Zhengzhou 450001, China\\%
$^{7}$Max Planck Institute for Astrophysics, Karl-Schwarzschild-Str. 1, D-85748, Garching, Germany\\%
$^{8}$School of Astronomy and Space Science, University of Chinese Academy of Sciences, Beijing 100049, China
}

\date{Accepted XXX. Received YYY; in original form ZZZ}

\pubyear{2025}

\begin{document}
\label{firstpage}
\pagerange{\pageref{firstpage}--\pageref{lastpage}}
\maketitle

\begin{abstract}
The dark matter halo mass function is one of the most fundamental predictions of structure formation theory and cosmological simulations. We present the full halo mass function in the $\Lambda$ cold dark matter ($\Lambda$CDM) model, ranging from a planetary mass   ($10^{-6}\,\mathrm{M}_\odot$; the thermal cutoff in the initial power spectrum for a fiducial  CDM particle mass of 100~GeV) to the mass of a rich galaxy cluster ($10^{15.5}\,\mathrm{M}_\odot$), and from redshift, $z=30$ to the present. To span this very large dynamic range, we combine our earlier  Voids-within-Voids-within-Voids (VVV) set of simulations (Wang et al) with large volume, lower resolution cosmological simulations. We develop a subsampling method to extract subvolumes from the original simulations, allowing us to reconstruct the global halo mass function from the biased underdense VVV regions. We show that the results agree reasonably well among the sets of simulations on  different scales and environments. We provide a fitting formula for the dark matter halo mass function based on the work of Reed et al. calibrated with our simulations, such that it can be applied at all scales, all environments and all times, with deviations of $\sim2-3\%$ at $z < 2$ and $\sim 7\%$ at higher redshift $z \gtrsim 5$. This formula is also accurate at least for a restricted set of 
models we tested with modest deviations from $\Lambda$CDM in the values of some of the cosmological parameters. 
A python code is publicly available at \href{https://github.com/haonan-zheng/hmfc}{https://github.com/haonan-zheng/hmfc}. 

\end{abstract}




\section{Introduction}
\label{sec:intro}

The mass function of dark matter haloes across different scales, environments and redshifts is a fundamental statistic and constraint on cosmic structure and associated galaxy formation theories \citep{Davis1985, White&Frenk1991}. 
Halo abundance bridges between theoretical dark matter physics and observed galaxy populations, serving as one of the primary inputs to methods such as abundance matching \citep{Frenk1988, Kravtsov2004, Vale2004, Behroozi2010, Moster2013} and halo occupation distribution fitting \citep{Benson2000, Jing1998, Peacock2000, Seljak2000} and, in return, constrains cosmological parameters \citep{Frenk1990, White1993, Henry2009, Allen2011ARA&A}. 
On the largest scales, the interpretation of ongoing and upcoming constraints from the abundance of galaxy clusters (e.g., weak lensing and X-ray/SZ surveys such as LSST and eROSITA, \citealt{Pillepich2012, Ivezic2019}) will require percent-level accuracy at the high mass end. 
On the smallest scales, a precise halo mass function down to planetary mass, is also crucial to forecast possible dark matter annihilation or decay signals, which are dominated by the smallest haloes owing to their large number and high concentration \citep[e.g.,][]{Bergstrom1999, Springel2008, Diemand2008, Grand2021, Zheng2024a, Delos2024}.  \cite{Delos2023} suggested that annihilation radiation is dominated by prompt cusps – small structures that formed ubiquitously as the very first generation of halo collapses.

At the faint end of the galaxy distribution, techniques such as strong gravitational lensing \citep[e.g.,][]{Dalal2002, Koopmans2005, Vegetti2010, Hezaveh2016, Li2016, He2022,Powell2025, Vegetti2026}, and gap detection in stellar streams \citep{Ibata2002, Johnston2002, Banik2021} are also pushing the frontier on the demographics of mini-haloes,  $M\lesssim 10^8\,\mathrm{M_\odot}$. 
At high redshifts, the halo mass function regulates the collapse of gas into the first stars and galaxies \citep[e.g.,][]{Barkana2001, Bromm2011}, thereby shaping the early UV galaxy luminosity function and the timing of reionization \citep[e.g.,][]{Furlanetto2006, Finkelstein2023,Lu2025}. 
These diverse applications strongly motivate the derivation of the halo mass function over the full dynamic range of mass and epochs. 

On the theoretical side, \citet{PS1974}, by connecting halo formation via spherical collapse with random walks in Gaussian random fields, provided a well-known analytical model for the halo mass function, the Press-Schechter formula (PS hereafter). 
This approach was further developed into the extended Press-Schechter model \citep[EPS hereafter, e.g.,][]{Bond1991, Bower1991, Lacey1993,Mo1996} by setting alternative starting points of the random walk, to predict the conditional halo mass function and mass accretion history in different environments \citep[][]{Gao2005, Faltenbacher2010, Zheng2024b, Liu2024, Delos2024b}. 
The accuracy of the formulae was  improved \citep[][i.e., the Sheth-Tormen model, ST hereafter]{ST1999, ST2002} to $\lesssim 10$ per cent level accuracy for $10^{10}-10^{14}\,\mathrm{M_\odot}$ haloes at $z < 8$ \citep{Reed2003}. 
Further efforts to generate empirical fits in cosmological simulations \citep[e.g.][]{Jenkins2001, Warren2006, Fernandez-Garcia2025} also reached a similar level of accuracy over different mass and redshift ranges. 

While early studies highlighted the universal form of the halo mass function at different redshifts, more recent studies such as those by \citet{Reed2007, Tinker2008, Crocce2010, Despali2016, Fiorilli2025, Benson2026}, using simulations of wider dynamical range, have reported small deviations from  universality and corrected them with time-evolving fitting parameters or variables. 
For example, \citet{Reed2007} provided two fits to the halo mass function: the first follows a universal form calibrated against the ST formula, while the second models halo abundance as a function of the effective slope of the power spectrum, thus implicitly incorporating time dependence to better match the simulation results. 

Due to the computational cost of simulations, these formulae have not been tested for small haloes ($\lesssim 10^5\,\mathrm{M_\odot}$) at $z \lesssim 10$, or across different environments \citep{Angulo2017}. 
This has only become possible with the Voids-within-Voids-within-Voids (VVV) simulations \citep{Wang2020}, which used a nested zoom-in strategy targeting voids, covering a wide mass range ($10^{-6}$–$10^{15.5}\,\mathrm{M_\odot}$) for one cold matter candidate, namely a 100 GeV WIMP (weakly interacting massive particle). 
\citet{Zheng2024b} used these simulations to test the EPS formula in the underdense volumes of the VVV simulations and found a deviation of $\gtrsim 20\%$, which worsens at intermediate redshifts ($z \sim 5$–10). 
These developments highlight the need for a halo mass function that covers a wide range of masses, environments, and redshifts, facilitating related theoretical studies as well as observational forecasts and comparisons. This is the motivation for the present work.

The paper is organized as follows. Section~\ref{sec:simulation} introduces the details of the simulations and fitting formulae being used or referenced. Our methods, 
results, discussions and conclusions are presented in Sections 
\ref{sec:nv}, \ref{sec:mass}, \ref{sec:discussions}, 
and \ref{sec:conclusions}, respectively. Additional tests of the halo definition, cosmological parameters, and subsampling method are presented in Appendices \ref{App-A}, \ref{App-B}, and \ref{App-C}, respectively. 

\section{Simulations and Methods}
\label{sec:simulation} 

\subsection{Simulations}

To fully cover such a wide halo mass range from a planetary mass to galaxy cluster mass, in this paper we use the VVV nested zoom N-body simulations \citep{Wang2020}, the P-Millennium N-body simulation \citep[][PMILL after]{Baugh2019}, and a bespoke N-body simulation VVV-2.8Gpc (VVV2.8 hereafter) with an even larger boxsize to increase the sample size at the high mass end and low redshift. 
Both the VVV and VVV2.8 are performed with the \textsc{Gadget-4} code \citep{Springel2020} and the PMILL is performed with a memory-efficient version of the \textsc{Gadget-3} code \citep{Springel2005}.
They all adopt the same cosmological parameters from \citet{Planck2014p16}: $\Omega_{\rm{m}} = 0.307$, $\Omega_{\Lambda} = 0.693$, $h = 0.6777$, $n_\mathrm{s} = 0.9611$ and $\sigma_8 = 0.8288$. 
For the the initial power spectrum, the \textsc{Camb} code \citep{Lewis2000} is used to calculate the large scale ($k \leq 7~\mathrm{Mpc}^{-1}$) part, which is smoothly joined to a BBKS fitting formula \citep[][with $\Gamma=0.1673$ and $\sigma_8=0.8811$]{BBKS1986} extrapolated at small scales ($k \geq 70~\mathrm{Mpc}^{-1}$); readers can refer to \citet{Wang2020} for details. 

The VVV simulations adopt a multi-zoom strategy to resolve extremely small structures at a reasonable computational cost \citep{Jenkins2010, Jenkins2013, Jenkins2013b}: an underdense and nearly spherical region is selected from the parent cosmological box (i.e., L0 for VVV), and resolved with higher resolution. This resimulated region is then used as the parent region for a further zoom-in. \citet{Wang2020} repeated this process and performed 8 levels of runs, labelled as L0-L7 from low to high resolution. 
Note that the VVV simulations adopted a particular dark matter model (i.e., 100 GeV WIMPs, \citealt{Wang2020}) that produced a small scale cut-off in the linear power spectrum for L7 and L8 -- but for the purposes of this paper we analyse a version of L7 where there is no resolved cut-off. 
Since PMILL has a similar boxsize but a 10 times better resolution than L0, we do not show the results of L0 in this paper and have checked that they are consistent with PMILL for well-resolved haloes. 
Following \citet{Wang2020} and \citet{Zheng2024b}, for L1-L7, we exclude the volume outside 0.8 times the radius of the high-resolution region from analysis, so as to avoid the numerical contamination from low-resolution particles. In Table \ref{tab:table1}, we list the simulation details of each run\footnote{Note that the rows L1-L7 of this table have been published in \citet{Zheng2024b}, we present them here again for the reader's convenience. } being used in this paper. 

In this work, only central haloes are considered, which are identified with the `friends-of-friends' \citep[\textsc{fof}, ][]{Davis1985} and \textsc{subfind} \citep{Springel2001, Dolag2009} algorithms, and the halo boundary is defined by the radius within which the enclosed density reaches 200 times the mean matter density of our Universe.

\begin{table}
 \caption{Properties of the different runs used in this work. Column 1: name of the run; column 2:
   size ($L_{\mathrm{box}}$ or $2\,r_{\mathrm{level}}$) of the analysed region 
   of each run at $z=0$. The entire box is used in VVV2.8 and PMILL, and in L1-L7 only the central sphere with a diameter $d_\mathrm{sphere} = 0.8\, d_\mathrm{region}$ is used -- $d_\mathrm{region}$ is the diameter of the entire high-resolution region; columns 3 and 4: mass and softening length of high-resolution particles; column 5: overdensity ($\delta_\mathrm{nl} = \bar{\rho} / \rho_{\mathrm{m}} - 1$) of analysed regions at $z=0$, $\rho_\mathrm{m}$ is the mean matter density of the universe. }
 \label{tab:table1}
 \begin{tabular}{clllc}
  \hline
  run &  size/$\mathrm{Mpc}$ & $m_{\mathrm{p}}$/$\mathrm{M_{\odot}}$ & $\epsilon$/$\mathrm{kpc}$ & $\delta_{\mathrm{nl}}$($z=0$) \\
  \hline

  VVV2.8 & $2.80 \times 10^{3}$  & $5.37 \times 10^{10}$  & $2.22 \times 10^{1}$ & $0.0$ \\    
  PMILL & $8.00 \times 10^{2}$  & $1.57 \times 10^{8}$  & $3.39 \ \times\ 10^0$ & $0.0$ \\
  L1 & $8.12 \times 10^{1}$  & $7.41 \times 10^{5}$  & $5.31 \times 10^{-1}$ & $-0.607$ \\ 
  L2 & $1.23 \times 10^{1}$  & $1.45 \times 10^{3}$  & $5.61 \times 10^{-2}$ & $-0.918$ \\ 
  L3 & $1.65 \ \times\ 10^{0}$  & $2.82 \ \times\ 10^{0}$  & $8.32 \times 10^{-3}$ & $-0.964$ \\ 
  L4 & $2.22 \times 10^{-1}$ & $5.50 \times 10^{-3}$ & $1.04 \times 10^{-3}$ & $-0.974$ \\ 
  L5 & $4.55 \times 10^{-2}$ & $5.75 \times 10^{-5}$ & $2.27 \times 10^{-4}$ & $-0.976$ \\ 
  L6 & $9.43 \times 10^{-3}$ & $2.60 \times 10^{-7}$ & $3.77 \times 10^{-5}$ & $-0.986$ \\ 
  L7 & $1.58 \times 10^{-3}$ & $8.55 \times 10^{-10}$& $5.28 \times 10^{-6}$ & $-0.984$ \\ 

  \hline
 \end{tabular}
\end{table}

\subsection{Formulae for halo mass functions}

\begin{table*}
 \caption{Minimum mass and $\ln \nu$ probed by P-Millennium (PMILL) and VVV-L7 (L7) at $z=0$, with and without our subsampling method that extracts subvolumes from the original simulation, as elaborated in Section \ref{sec:subsample}. Column 1: name of the (subsampled) run, with `S' denoting a smaller subsampled volume and `L' a larger one; column 2: size ($L_\mathrm{box}$ for PMILL, $2\;\!r_\mathrm{level}$ for L7, $2\;\!r_\mathrm{region}$ for L7 subsampled) of the analysed region; column 3: mass of the smallest haloes probed, taken as the mass of 50 dark matter particles; column 4: overdensity of the (subsampled) region; columns 5 and 6: corresponding linear overdensity and cosmic variance; column 7: minimum $\ln \nu$ corresponding to $M_\mathrm{min}$; column 8: number of obtained subsampled regions whose deviation from the target overdensity is smaller than 0.001; column 9: summed weights of subsampled regions, indicating the number of independent regions, see Section \ref{sec:subsample} for details. }
 \label{tab:table2}
 
 \begin{tabular}{ccccccccc}
  \hline
  run & size/$\mathrm{Mpc}$ & $M_\mathrm{min}/\mathrm{M}_\odot$ & $\delta_{\mathrm{nl}}$ & $\delta_0$ & $\sigma^2_0$ & $\ln \nu_\mathrm{min}$ & $N_\mathrm{realization}$ & $\sum\omega_\mathrm{region}$ \\
  \hline
  
  PMILL & $8.00 \times 10^{2}$ & $7.83 \times 10^9$ & $0$ & $0$ & $1.13 \times 10^{-4}$ & $-0.87$ & 1 & – \\ \hline
  L7 & $1.58 \times 10^{-3}$ & $4.28 \times 10^{-8}$ & $-0.984$ & $-20.51$ & $2.35 \times 10^2$ & $0.10$ & 1 & – \\ \hline
  \multirow{6}{*}{L7 subsampled (S)} & \multirow{6}{*}{$2.83 \times 10^{-5}$} & \multirow{6}{*}{$4.28 \times 10^{-8}$} & 0 & $0$ & $3.95 \times 10^2$ & $-2.23$ & 331 & 160.74 \\ 
  & & & $-0.2$ & $-0.24$ & $4.00 \times 10^2$ & $-2.09$ & $449$ & 226.98\\ 
  & & & $-0.4$ & $-0.60$ & $4.06 \times 10^2$ & $-1.90$ & $926$ & 361.31\\ 
  & & & $-0.6$ & $-1.25$ & $4.16 \times 10^2$ & $-1.64$ & $2196$ & 752.03\\ 
  & & & $-0.8$ & $-2.79$ & $4.33 \times 10^2$ & $-1.17$ & $8783$ & 2095.19\\  
  & & & $-0.9$ & $-5.20$ & $4.50 \times 10^2$ & $-0.70$ & $35154$ & 5827.73\\ \hline
  \multirow{6}{*}{L7 subsampled (L)} & \multirow{6}{*}{$4.00 \times 10^{-5}$} & \multirow{6}{*}{$4.28 \times 10^{-8}$} & 0 & 0 & $3.71 \times 10^2$ & $-2.28$ & $285$ & 88.94 \\ 
  & & & $-0.2$ & $-0.24$ & $3.76 \times 10^2$ & $-2.14$ & $372$ & 105.40\\ 
  & & & $-0.4$ & $-0.60$ & $3.82 \times 10^2$ & $-1.95$ & $763$ & 172.11\\ 
  & & & $-0.6$ & $-1.24$ & $3.92 \times 10^2$ & $-1.69$ & $1567$ & 290.77\\ 
  & & & $-0.8$ & $-2.79$ & $4.08 \times 10^2$ & $-1.23$ & $6960$ & 820.20\\ 
  & & & $-0.9$ & $-5.20$ & $4.24 \times 10^2$ & $-0.76$ & $27559$ & 2185.25\\ 

  \hline
 \end{tabular}
 
\end{table*}

Since L1-L7 runs focus on extremely underdense regions, the halo number densities inside are inevitably much lower than the global average value in the whole universe, and therefore cannot be directly used for fitting. 
However, as indicated by the results of \citet{Zheng2024b}, an approach exists based on the extended Press-Schechter theory \citep{Bond1991, Lacey1993, Mo1996} and the assumption of independent random walks, to bridge the local and global halo mass function, through a transformation of measurement space: 
\begin{equation}
    \nu = \frac{\delta_1 - \delta_0}{\sqrt{\sigma^2_1 - \sigma^2_0}},
\end{equation}
where $\delta_1=\delta_\mathrm{c}/D(z)$, $\delta_\mathrm{c}=1.68647$, $D(z)$ is the growth factor at redshift $z$, and $\delta_0$ is the linear overdensity of the measured region \citep[see Eq. 6 of][for the conversion from the non-linear overdensity $\delta_\mathrm{nl}$]{Zheng2024b}, and $\sigma^2_1$ and $\sigma^2_0$ denote the cosmic variance $\sigma^{2}(M)$ smoothed in a tophat filter containing masses corresponding to the halo mass $M_1$ and region mass $M_0$, respectively. Note that in the whole universe, $\delta_0$ and $\sigma_0$ are both zero, and $\nu$ returns to $\delta_1/\sigma_1$. With this method, the formulae of halo mass function from PS can be written in this equation: 
\begin{equation}
    \label{eq:EPS}
    f_\mathrm{PS}(\nu) = \sqrt{\frac{2}{\pi}}~\nu~\exp \left({-\frac{\nu^2}{2}}\right), 
\end{equation}
where $f$ is the mass function defined as the fraction of mass inside virialized haloes per unit interval of $\mathrm{d}\ln\nu$ following \citet{Reed2007}, and can be converted into the differential halo mass function with: \begin{equation}
    \frac{\mathrm{d}n}{\mathrm{d}M} = \frac{\rho_\mathrm{m}}{M} \frac{\mathrm{d}\ln \nu}{\mathrm{d}M}f(\nu). 
\end{equation} 

In this approach, the ST and R07 mass functions \citep{ST1999, ST2002, Reed2007} can be transformed as modified versions of Eq.~(\ref{eq:EPS}):
\begin{equation}
    f_\mathrm{ST}(\nu) = A_\mathrm{ST}\sqrt{\frac{2a_\mathrm{ST}}{\pi}}\left(1 + a_\mathrm{ST}^{-p_\mathrm{ST}}\nu^{-2p_\mathrm{ST}}\right)~\nu~\exp \left({-\frac{1}{2}a_\mathrm{ST}\nu^2}\right), 
\end{equation} 
\vspace{-0.3cm}
\begin{equation}
    \begin{split}
    f_\mathrm{R07}(\nu) = & A_\mathrm{ST}\sqrt{\frac{2a_\mathrm{ST}}{\pi}}\left(1 + a_\mathrm{ST}^{-p_\mathrm{ST}}\nu^{-2p_\mathrm{ST}}+0.2G_1\right)\times \\
    & ~\nu~\exp \left({-\frac{1}{2}c_\mathrm{R07}a_\mathrm{ST}\nu^2}\right).
    \end{split}
\end{equation} 
Here, $A_\mathrm{ST}=0.3222$, $a_\mathrm{ST}=0.707$, $p_\mathrm{ST}=0.3$, $c_\mathrm{R07}=1.08$, and $G_1$ is a Gaussian correction term defined as: 
\begin{equation}
    G_1 = \exp\left\{-\frac{\left[\;\!\ln (\nu / \delta_\mathrm{c})-0.4\;\! \right]^2}{2(0.6)^2} \right\}.
\end{equation} 

\subsection{The methods of subsampling}
\label{sec:subsample}

\begin{figure*}
    \centering
    \includegraphics[width=2\columnwidth]{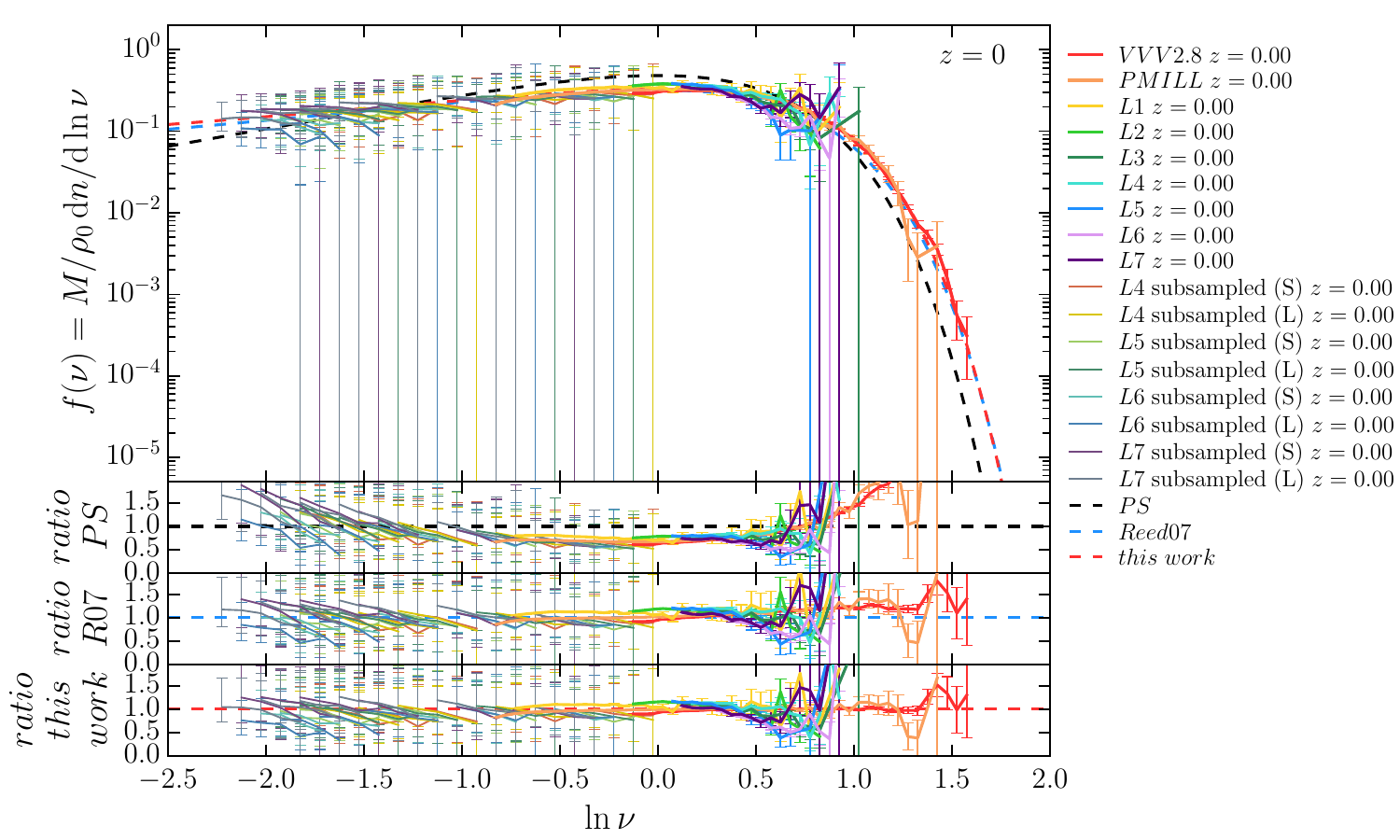}
    \includegraphics[width=2\columnwidth]{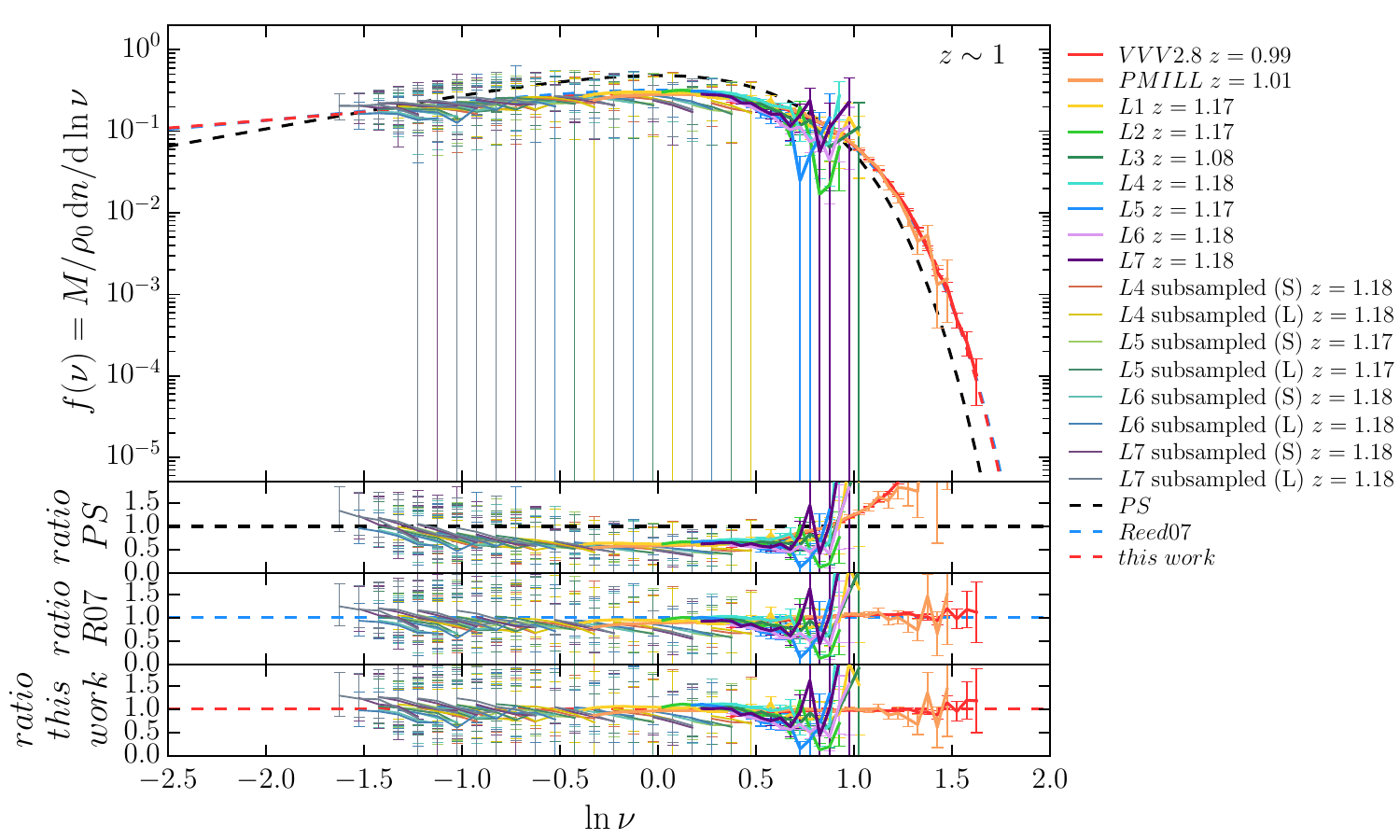}
    
    \caption{Halo mass functions at $z=0$ and $z \sim 1$. In the top panels, thick solid lines show the halo mass functions with Poisson errors from the original simulation volumes (i.e., VVV2.8, PMILL, L1-L7), while the thin solid lines show those from subvolumes subsampled from L4-L7 with different local densities (see Section \ref{sec:subsample} for details) with the 16th-84th percentiles shown. The dashed lines show the predictions of different fitting formulae. The bottom panels show the ratios of simulations to analytical predictions. The R07 formula shows good agreement with the simulations at $z = 0$ and $z \sim 1$, with a slight high-$\nu$ deviation at $z=0$ that is corrected by our new formula. }

    \vspace{-0.30cm}
    \label{fig:fig1}
\end{figure*}

Since we aim to resolve and fit the halo mass function at scales over as wide a range as possible in the $f$ - $\nu$ space, it becomes crucial to extend the coverage of simulations in $\nu$. 
Even though our simulations resolve haloes down to planetary mass, as $\nu^2$ is defined as $(\delta_1 - \delta_0)^2 / (\sigma^2_1 - \sigma^2_0)$, the extension to the low mass end can be significantly limited by the extreme underdensity of the VVV high level volumes taken as a whole especially at late times. 
We list the overdensity and the corresponding linear overdensity for the VVV L7 region at the top of Table \ref{tab:table2}, together with the lower limit in $\ln \nu$ that can be reached for the entire L7 region ($\ln \nu=0.10$) or from the P-Millennium ($\ln \nu=-0.87$). 

Therefore, we adopt a subsampling method that extracts smaller and denser subvolumes from each level, to obtain the $f(\nu)$ function at the low mass end. 
Each subsampled region is treated as an individual region when calculating the halo mass function, with $\delta_0$ and the cosmic variance $\sigma_0^2$ determined by its own overdensity and size. 
Haloes in a subsampled region are then assigned $\nu$ according to the properties of that subregion rather than the entire original VVV level. 
The same halo may thus be included in multiple subsampled regions and assigned different values of $\nu$, enabling us to extend our measurement to the low $\nu$ regime. 
This method assumes that, once the properties of a subsampled region are specified, the halo formation inside is largely independent of the larger scale VVV environment, as motivated by EPS theory and further tested in Appendix \ref{App-C}.

The size of the subsampled spherical region is determined both by the level volume and by the resolution: 
\begin{equation}
    r_\mathrm{region} = \sqrt{r_\mathrm{level}~r_\mathrm{typical\ halo}}~, 
\end{equation}
here, $r_\mathrm{level}$ is the radius of the analysed region of each level, $r_\mathrm{typical\ halo}$ is the virial radius of a `typical' halo ($M_{200}=625\,m_\mathrm{dm}$, and $5000\,m_\mathrm{dm}$, marked as `S' and `L' volumes respectively since they correspond to smaller and larger volume sizes), so that a region with a particular overdensity is neither too rare to find, nor too small to contain a sufficient number of haloes.

In practice, we use the global optimization algorithm implemented in the \textsc{python} package \textsc{scipy.shgo} to search for these regions, using the deviation from the required density as an optimizing function of the coordinates of the region centre, i.e.,
\begin{equation}
    \Delta(\boldsymbol{x}) = \lvert \delta_\mathrm{nl}(\boldsymbol{x}, r_\mathrm{region}) - \delta_\mathrm{nl,\,required} \rvert, 
\end{equation}
here, $\Delta$ is the function to be minimized, $\boldsymbol{x}$ is the coordinate of the region centre acting as a variable, and $\delta_\mathrm{nl,\,required}$ is the desired non-linear density. 
We manage to find hundreds to ten thousands of the corresponding regions with overdensity deviation $\Delta < 0.001$ for $\delta_\mathrm{nl,\,required} = 0, -0.2, -0.4, -0.6, -0.8, -0.9$, with the number of regions varying for different overdensities, sizes, levels, and redshifts. 

For example, extracting subvolumes with cosmic mean density (i.e., $\delta_\mathrm{nl}=\delta_0=0$) and a radius of 20 pc inside L7, enables us to reach $\ln \nu=-2.28$, compared to $\ln \nu=0.10$ of the original L7 volume The minimum $\ln\nu$ achieved with other subsampled realizations of L7 are listed in Table~\ref{tab:table2}. 
An additional benefit of this method is that it selects multiple realizations instead of a single realization of each level (285 realizations, compared to one realization from the whole L7 volume in the aforementioned case), thereby capturing cosmic variance, even with the same volume size and overdensity. 

Since the halo mass function can vary among these regions due to the cosmic variance, we adopt the median value for each bin as combined from each subvolume to generate our fits. 
It is worth noting that our region-finding procedure does not prevent overlap among the subsampled spherical regions. To account for haloes double-counted in overlapping subvolumes, it is required to assign each region a weight $\omega_{\mathrm{region}}$ when computing the median.\footnote{For ordered values $f_1$, $f_2$, $\dots$, $f_n$ with corresponding weights $\omega_1$, $\omega_2$, $\dots$, $\omega_n$, the weighted median is defined as the value $f_k$ at which the cumulative weight first exceeds $0.5 \sum_{i=1}^{n} \omega_i$.} 
For each selected region $i$, we calculate its overlapping volume with other regions $V_{\mathrm{overlap}\ i} \equiv \sum_{j \neq i} V_{\mathrm{region} \ j\ \cap\ \mathrm{region}\ i}$ and then determine its weight $\omega_{\mathrm{region}\ i}$ as:
\begin{equation}
    \omega_{\mathrm{region}\ i} = V_{\mathrm{region}\ i}/(V_{\mathrm{region}\ i} + V_{\mathrm{overlap}\ i}). 
\end{equation}
The greater the overlap, the less representative a subvolume would be considered: for example, two highly overlapped spheres will have $\omega_\mathrm{region}=0.5$, which is equivalent to one of such sphere. 

In the upper panel of Fig. \ref{fig:fig1}, we show
the halo mass functions from subsampling: the thin solid lines show the weighted median values, and the corresponding error bars represent the weighted 16-84th scatter. 
We find the scatter is quite large, with an amplitude of $\sim$ 0.3 dex as shown by error bars, with the `S' samples generally showing larger scatter compared to `L', likely due to their relatively small volume sizes. 
However, their median values (i.e., thin solid lines) remarkably converge with each other, despite the contrast environmental density, and at $\ln \nu \in [-0.75, 0]$ they also converge well with those obtained from the original simulation volumes (thick orange line, i.e., PMILL), with deviations mostly within $\sim 0.05$ dex, confirming that our method is able to reproduce the mean halo mass function in the whole universe via this EPS-motivated subsampling method.

\begin{figure*}
    \centering
    \includegraphics[width=2\columnwidth]{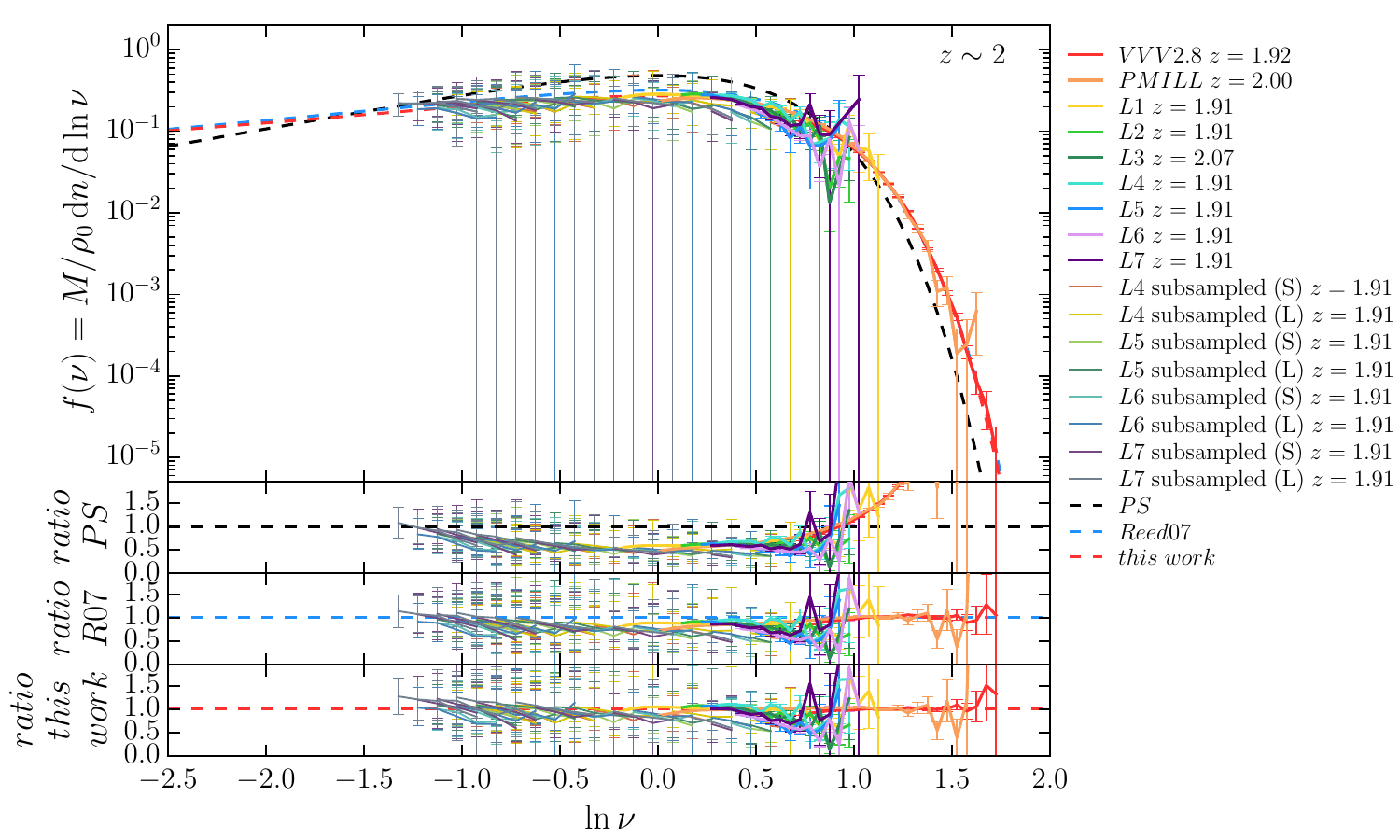}
    \includegraphics[width=2\columnwidth]{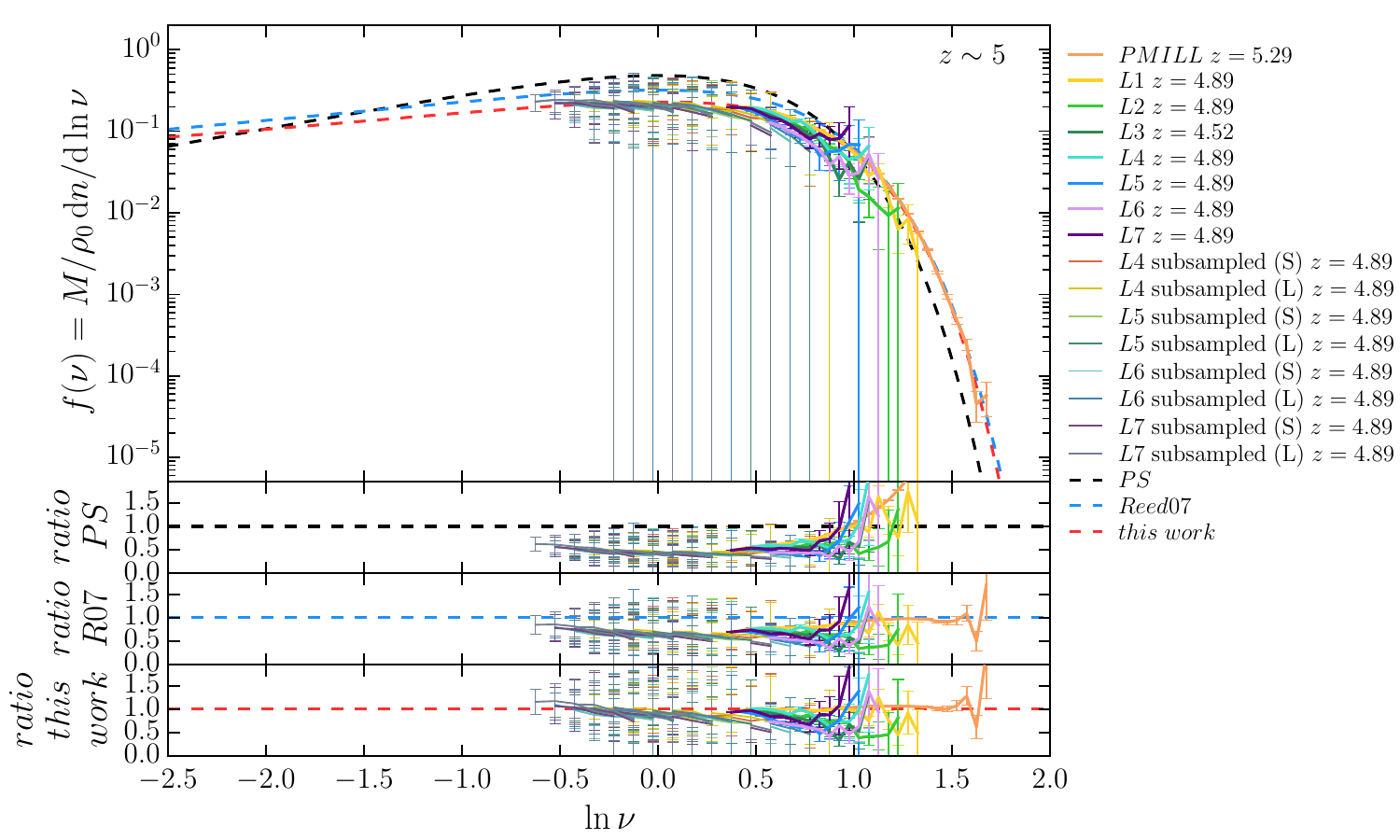}
    \caption{Same as Fig.~\ref{fig:fig1}, but for $z \sim 2$ and 5 (VVV2.8 is not shown at $z \gtrsim 5$ due to having too few haloes). At the high mass end, PS significantly underestimates the halo abundance, while both R07 and our formula show good agreement with the simulations. At the low mass end, R07 shows a deviation increasing with redshift, which is calibrated away in our formula. }

    \label{fig:fig2}
\end{figure*}

\begin{figure*}
    \centering
    \includegraphics[width=2\columnwidth]{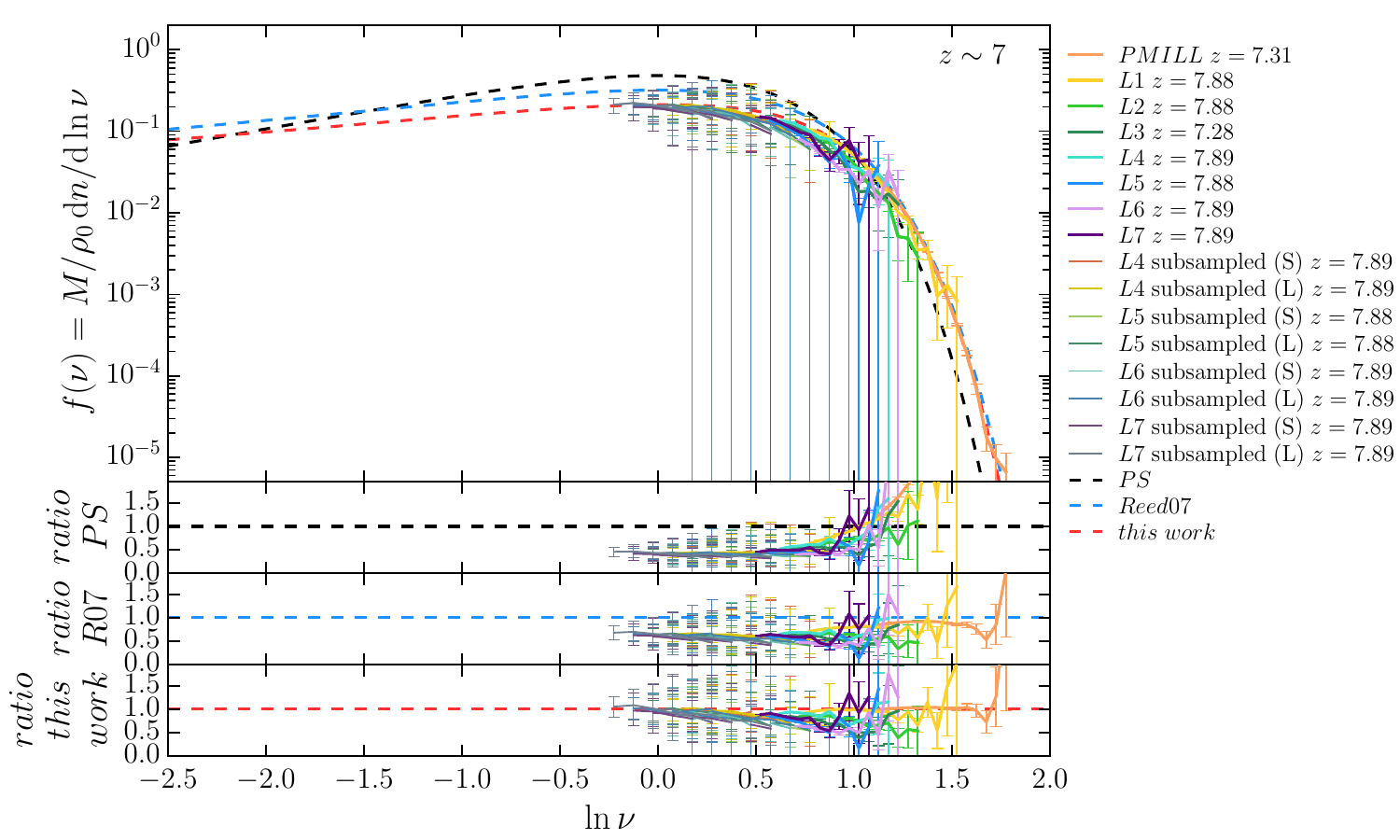}
    \includegraphics[width=2\columnwidth]{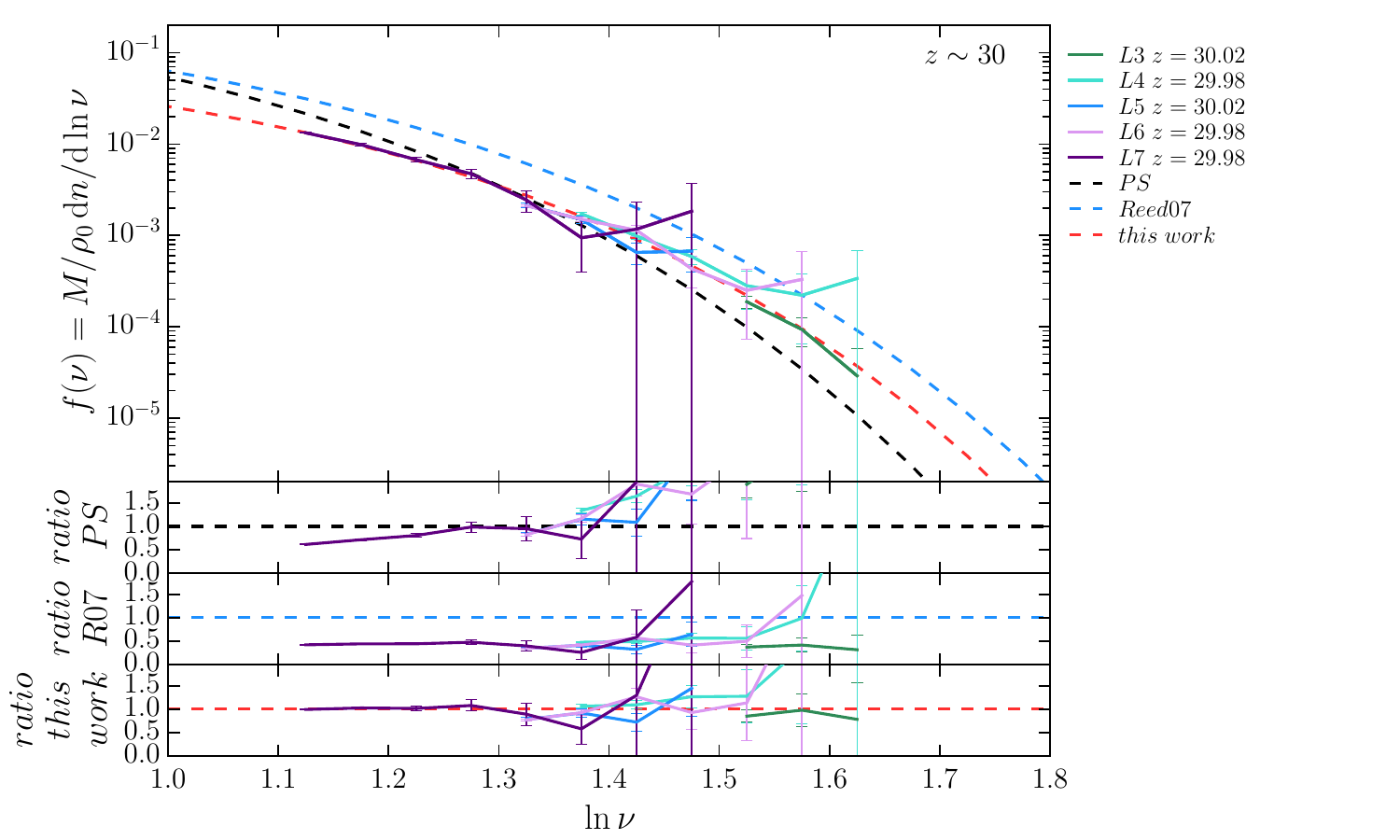}
    \caption{Same as Fig.~\ref{fig:fig1}, but for $z \sim 7.5$ and 30 (subsampling is not implemented at $z \sim 30$ as there are too few haloes; PMILL, L1, and L2 are not shown for the same reason). The deviation of both the PS and R07 formulae relative to the simulations grows at higher redshifts compared to lower ones, whereas our formula reduces the discrepancy from a factor of $\sim 2$ to $\sim 7\%$. }

    \label{fig:fig3}
\end{figure*}

\section{Halo mass function in the $f-\nu$ space}
\label{sec:nv}

In Fig.~\ref{fig:fig1}, \ref{fig:fig2}, and \ref{fig:fig3}, we show the halo mass function in the $f(\nu)\ -\ \ln \nu$ space at redshifts $z \sim 0$, 1, 2, 5, 7, and 30. 
Thick solid lines represent the halo mass functions in the original simulation volumes, while the thin solid lines represent the halo mass function obtained via `subsampling' as described in Section \ref{sec:subsample}.

\subsection{Comparison with the PS and Reed2007 formula}

In Fig.~\ref{fig:fig1}, the dashed black lines represent the prediction of the PS theory. 
We find the PS formula generally overestimates the halo abundance at the low mass end particularly for $\ln \nu \in [-1.2, 1]$, but transitions to underestimation at $\ln \nu \gtrsim 1$. 
This result reflects to the known inaccuracy of PS in fitting simulation results due to the over-simplified spherical collapse model, which has been noted by many studies in the literature \citep[e.g.,][]{ST1999, ST2002, Jenkins2001, Reed2003, Reed2007, Zheng2024b}. 

To quantify the deviation of formulae from simulation results, we define the deviation $\varepsilon$ as follows: 
\begin{equation}
    \label{single}
    1 + \varepsilon_\mathrm{single\ bin}\, = \, \langle ratio\rangle_\mathrm{within\ single\ bin}, 
\end{equation}
\vspace{-0.5cm}
\begin{equation}
\label{epsilon}
    1 + \varepsilon\, = \, \langle 1 + \varepsilon_\mathrm{single\ bin}\rangle_\mathrm{different\ bins}, 
\end{equation}
here, $\langle \ \rangle$ denotes the average at log scale while taking the absolute value (i.e., $\exp\left(\dfrac1n \sum_{}^n|\log x_i|\right)$), the $ratio$ is the ratio between simulation and formula prediction, and $\varepsilon_\mathrm{single\ bin}$ denotes the deviation in one single bin, averaged among different realizations (i.e., different runs and subsampling).\footnote{Following \citet{Zheng2024b}, we exclude the bins with particle number or halo number below 50 due to resolution considerations. }
The deviation is measured at log-scale, so that overestimation and underestimation can be considered equally. 
With this approach, we count the contribution of each bin equally, rather than over-counting certain bins with more realizations. The deviation here should be decomposed into the intrinsic discrepancy among simulation runs, $\varepsilon_\mathrm{simulation}$, and the actual deviation of the fits to simulations, $\varepsilon_\mathrm{fit}$. $\varepsilon_\mathrm{simulation}$ can be calculated by replacing the $ratio$ in eq. \ref{single} with the ratio of each simulation run to the geometric mean value in each bin, corresponding to the minimum $\varepsilon$ a fit can achieve; then $\varepsilon_\mathrm{fit}$ is calculated as $\varepsilon_\mathrm{fit}\, = \varepsilon - \varepsilon_\mathrm{simulation}$. 

Thereby, we confirm the inaccuracy of the PS, with a deviation $\varepsilon_\mathrm{fit,\,PS} = 34.05\%$ at $z=0$ as listed in Table \ref{tab:table3}. 
We then extend the analysis to higher redshifts in Fig. \ref{fig:fig2} and \ref{fig:fig3}, finding that the deviation increases and reaches the highest value at $z \sim 7$, but it provides a better fit to the simulations at $z=30$, aligning with our previous findings in \citet{Zheng2024b}. 

\begin{table}
 \caption{The deviation (in percent) of different halo mass function formulae compared to simulation results, $\varepsilon_\mathrm{fit}$, as a function of redshift. $\varepsilon_\mathrm{simulation}$ are listed as an estimation of the discrepancy among simulation runs as a reference, mostly contributed by the numerical noise in relatively poorly resolved bins, see main text for details. }
 \label{tab:table3}
 \begin{tabular}{ccccccc}
  \hline
  $z$ & 0 & 1 & 2 & 5 & 7 & 30 \\

  \hline

  $\varepsilon_\mathrm{fit,\,PS}$ & 34.05 & 52.46 & 65.72 & 90.28 & 114.56 & 31.09 \\
  $\varepsilon_\mathrm{fit,\,R07}$ & 6.27 & 6.21 & 10.32 & 25.91 & 39.67 & 125.08 \\
  $\varepsilon_\mathrm{fit,\,this\ work}$ & 1.89 & 3.08 & 3.03 & 6.28 & 7.58 & 7.11 \\

  \hline
    
  $\varepsilon_\mathrm{simulation}$ & 8.37 & 6.97 & 7.02 & 7.60 & 8.39 & 2.03 \\

  \hline
 \end{tabular}
\end{table}

We further test the \citet{Reed2007} formula and plot it with blue dashed lines.\footnote{We also tested another formula \citet{Reed2007} provided (i.e., their Eq. 11 and 12), with the effective slope of power spectrum $n_\mathrm{eff}$ taken into account. 
We note that the extension into the low-mass regime drives $n_\mathrm{eff}$ toward $-3$, which leads to a divergence in the formula and a prediction that deviates strongly from simulations. Thus in this work, we only adopt their first formula independent of $n_\mathrm{eff}$. }
We find the \citet{Reed2007} formula shows good agreement at the low $\nu$ end, while underestimates by $\sim 10\%$ at $\ln \nu \gtrsim 0.8$. 
This result aligns with the left panel of Figure 6 in \citet{Reed2007}, where simulation results at low redshifts are higher than this fit in a similar mass range, but with an even larger deviation. 
We note that this is due to the different mass definition, as \citet{Reed2007} adopted halo FOF mass and we use $M_{200}$ in this paper -- we checked our results using the halo FOF mass (not shown) and did observe a larger discrepancy between the simulations and the formula prediction. 
At higher redshift, this deviation at the high $\nu$ end disappears by $z \sim 1$, but an overestimation appears at the low $\nu$ end and gradually increases to $\sim 30\%$ at $z=7.89$ and $50\%$ at $z=30$. 
The evolution of $\varepsilon_\mathrm{fit,\, R07}$ also shows an increasing discrepancy when extended to high redshift: it suggests a deviation of $\sim 6\%$ at $z=0$ and $1$, but rises to 39.67\% and 125.08\% at $z=7$ and $30$, respectively. 

This suggests a time evolution of the halo mass function $f(\nu)$, which has likewise been suggested by many commonly used fitting formulae \citep[e.g.,][]{Tinker2008, Crocce2010}. 
Taking the low $\nu$ end as an example, the \citet{Reed2007} formula already provides a good fit at $z=0$, but increasingly overestimates the halo abundance at higher redshift, reaching a factor of $\sim 1.5$ at $z = 7$. 
Therefore, simply re-fitting the parameters with a single set of constant values cannot simultaneously account for the halo abundance across the full redshift range we aim to cover, meaning that improving the agreement at high redshift would compromise the good performance at low redshift. 
This leads to the introduction of explicit redshift dependence into the parameters, and hence into the fitting formula.

\subsection{Empirical correction to the Reed2007 formula}\label{Empirical}

Motivated by the observed evolution of the deviations, we propose an empirical correction to the \citet{Reed2007} formula: 
\begin{equation}
    \begin{split}
    f_\mathrm{this\ work}(\nu) = & A_\mathrm{ST}\sqrt{\frac{2a_\mathrm{ST}}{\pi}}\left[1\! + \!\left(a_\mathrm{ST}v^2\right)^{-p_\mathrm{this\ work}}\!\!+\!0.2\;\!G_\mathrm{this\ work}(\nu,z)\right]\\
    & \times \Gamma_\mathrm{this\ work}(\nu, z) ~\nu~\exp \left({-\frac{1}{2}c_\mathrm{this\ work} a_\mathrm{ST}\nu^2}\right),
    \end{split}
\end{equation} 
where $p_\mathrm{this\ work}(z)$ = 0.33, $c_\mathrm{this\ work}$ = 1.04, 
\begin{equation}
    \begin{split}
    G_\mathrm{this\ work}(\nu, z) = \, & \mathrm{min}(2.70,\, 2.35-0.36z+0.07z^2) \times \\ &
    \exp\left\{-\frac{\left[\;\!\ln (\nu / \delta_\mathrm{c})-\min(0.5 + 0.05z, 0.6)\;\! \right]^2}{2(0.55)^2} \right\}, 
    \end{split}
\end{equation}
and $\Gamma_\mathrm{this\ work}(\nu, z)$ is a correction term defined as follows: 
\begin{equation}
    \Gamma_\mathrm{this\ work}(\nu, z) = \frac12 \alpha(z) \left\{1 - \tanh \left[\;\!\beta(z)\left(\ln \nu - 1.78 \right) \right] \right\}, 
\end{equation} 
\vspace{-0.5cm}
\begin{equation}
    \alpha(z) = 0.252 + 1.53\;\! \theta(z) - 2.028\;\! \theta^2(z) + 1.382\;\! \theta^3(z), 
\end{equation} 
\vspace{-0.5cm}
\begin{equation}
    \beta(z) = 4.10 - 2.10\;\! \theta(z), 
\end{equation} 
\vspace{-0.5cm}
\begin{equation}
    \theta(z) = \exp \left(- \frac{1+z}{10} \right), 
\end{equation} 
$G_\mathrm{this\ work}(\nu, z)$ and $c_\mathrm{this\ work}$ account for the correction to the high mass end of the halo mass function, while $\Gamma(\nu, z)$ suppresses the low mass end of the halo mass function with a hyperbolic tangent function, yielding a constant reduction to a factor of $\alpha(z)$ as $\ln \nu \rightarrow -\infty$, but approaching 0 at the high $\nu$ end to balance the smaller $c$ compared to R07, as designed to fit the deviation from the R07 formula to simulation results. 
We thereby correct the original R07 formula at the low mass end (and the high mass end at $z<1$), while preserving agreement at intermediate masses. 

In the bottom rows of Fig.~\ref{fig:fig1}, \ref{fig:fig2}, and \ref{fig:fig3}, we show the ratio between simulation results and our fitting formula, and find that it mostly overlaps with the horizontal dashed line, especially at the low mass end at high redshift $z$, as corroborated by the average deviation $\varepsilon_\mathrm{fit}$ at a $\sim 2-7\%$ level in Table~\ref{tab:table3}, highlighting the great performance of our fit across such a broad range of $\ln \nu \in [-2, 1.8]$ and $z \in [0, 30]$. 

\section{Halo mass function in the $f-M$ space}
\label{sec:mass}

In the previous section, we provided a fitting formula for the mass function $f(\nu)$ as a function of $\nu$, but to calculate the actual halo number density $n$ at a given mass $M$, one would need to convert $\nu$ in the formula as follows: 
\begin{equation}
    \nu = \dfrac{\delta_\mathrm{c}}{D(z)\sigma(M)}, 
\end{equation} 
\vspace{-0.2cm}
\begin{equation}
    \sigma^2(M) = \frac{1}{2 \pi^2} \int^{\infty}_{0} k^2 P(k) W^2(k;M) \mathrm{d}k, 
\end{equation} 
here $P(k)$ is the linear power spectrum of the density fluctuations rescaled at $z=0$, and $W(k;M)$ is a real-space top-hat filter \citep[e.g.][]{Lacey1993, Mo2002}. 

For the convenience of readers, we provide a fit for this conversion in the \citet{Planck2014p16} cosmology, which is precise to 0.1\% level for $M\in \left[10^{-6}, 10^{16} \right]\,\mathrm{M_\odot}$: 
\begin{equation}
    \begin{split}
    \ln \sigma(M) = & -1.2723^{\log M - 11.79} - 0.0665 \log M + 2.6323 \\
     & + 0.0177\mathcal{N}(14.60, 4.4) +  0.0432\mathcal{N}(6.15, 21) \\
     & + 0.1508\mathcal{N}(-1.59, 34) +  0.0270\mathcal{N}(-7.81, 12), \\
    \end{split}
\end{equation} 
where the $\mathcal{N}(\mu, \varsigma^2) \equiv \exp{\left[-\dfrac{(\log M - \mu)^2}{2\varsigma^2} \right]}$ are correction terms with the Gaussian distribution function. 

In a flat universe with dark energy as a cosmological constant, $D(z) \equiv g(z)/\left[g(0)(1+z)\right]$ can be expressed in terms of Gaussian hypergeometric functions \citep{Chernin2003}, with 
\begin{equation}
    g(z) = {}_2F_1\left[\frac13, 1; \frac{11}6; -\frac{\Omega_\Lambda}{\Omega_\mathrm{m}}(1+z)^3 \right], 
\end{equation} 
while for a flat universe with dark energy being described with a time-varying equation of state (also known as `$w_0w_a$' model, i.e., the equation of state ratio follows $w(z)=w_0+w_a\, z/(1+z)$), \citet{Linder2005} provided an analytical fit accurate to a sub-percent level:
\begin{equation}
    g(a) = \exp\left\{{\int^a_0 \mathrm{d} \ln a \left[\Omega_\mathrm{m}(a)^\gamma -1 \right]} \right\}, 
\end{equation} 
where $\Omega_\mathrm{m}(a) = \Omega_\mathrm{m} a^{-3} / \left[H(a)/H_0\right]^2$, and
\begin{equation}
    \gamma = \left\{
    \begin{array}{l}
        \!0.55 + 0.05 \left[1 + w(z=1) \right]\ \ \ \  (w > -1) \\
        \!0.55 + 0.02 \left[1 + w(z=1) \right]\ \ \ \  (w < -1) \\
    \end{array} 
    \right.  \ .
\end{equation} 

\begin{figure*}
    \centering
    \includegraphics[width=1.0\columnwidth]{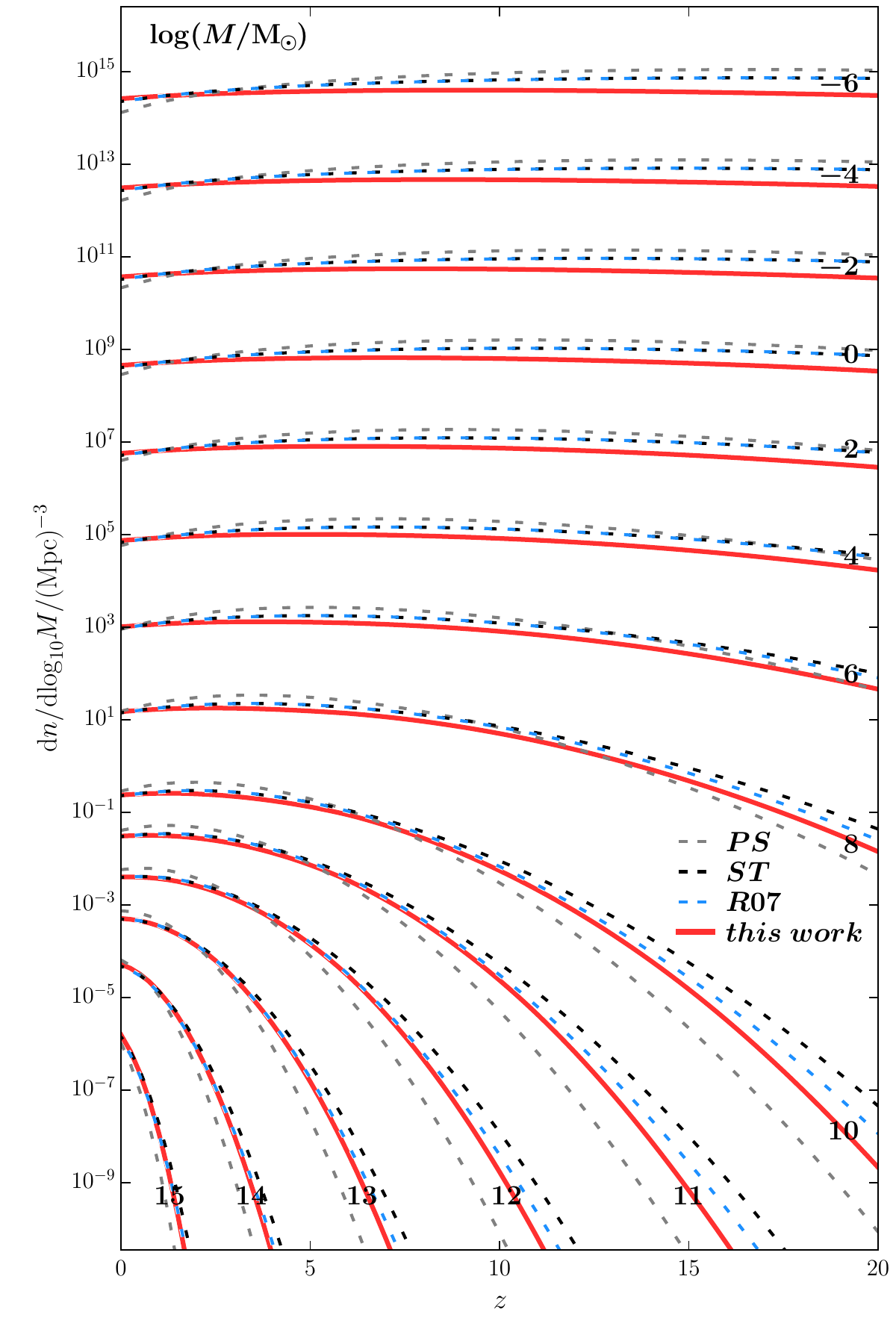}
    \includegraphics[width=1.0\columnwidth]{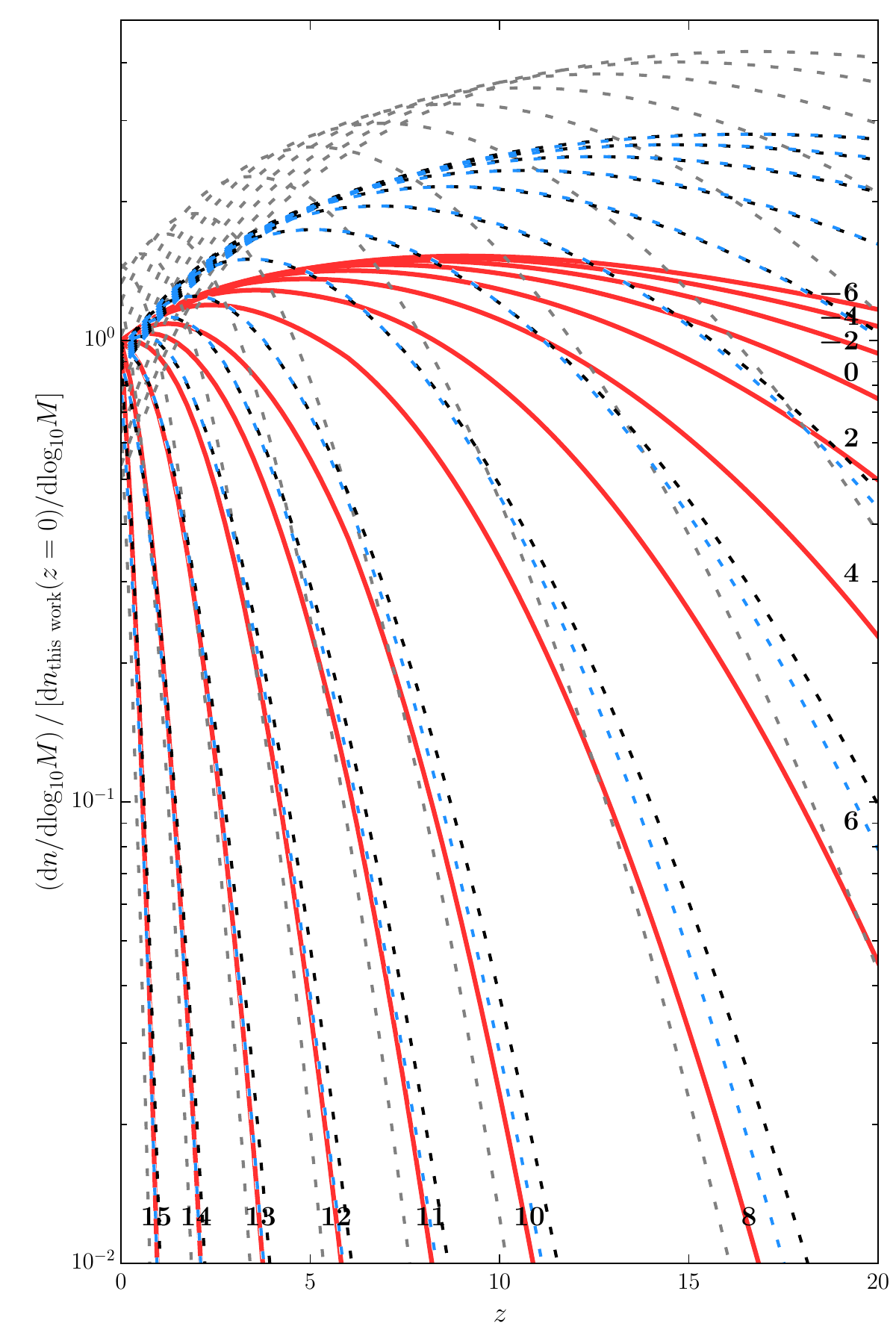}
    \vspace{-0.30cm}
    \caption{Left: The evolution of the halo mass function for different masses ($10^{-6}-10^{15}\,\mathrm{M_\odot}$, as represented by each set of curves labelled with the corresponding $\log (M/\mathrm{M_\odot)}$) in the whole universe, predicted by different formulae (gray: PS, black: ST, blue: R07, red: this work); right: same as the left panel, but rescaled by the halo abundance our fit predicts at present for a closer look at the difference. Starting with the modification of R07 over ST at the high mass end ($M \gtrsim 10^8\,\mathrm{M_\odot}$), we have further calibrated the fit at the low mass end ($M \lesssim 10^{10}\,\mathrm{M_\odot}$) and especially at high redshift; our slight calibration at the highest $\nu$ end becomes only significant at the lowest redshift ($z<1$) and for $M \gtrsim 10^{14}\,\mathrm{M_\odot}$.}
    \label{fig:fig4}
\end{figure*}

In Fig.~\ref{fig:fig4}, we show the predicted halo mass function in the whole $\Lambda$CDM universe, over a wide range of halo mass and redshift ($10^{-6}-10^{15}\,\mathrm{M_\odot}$, $0<z<20$). 
Consistent with our previous figures and the literature, compared to the assumption of spherical collapse with PS, the spheroidal collapse with ST converts more mass into massive haloes while suppressing the abundance of haloes with lower mass ($\sim 10^{10}\,\mathrm{M}_\odot$, depending on $z$). 
The R07 formula mainly steepens the slope (in the $f-\nu$ and $n-M$ space) at the high mass end, predicting a lower halo mass function at $M \gtrsim 10^8\,\mathrm{M_\odot}$. 
Our fit further calibrates the R07 description at the lower mass end ($M \lesssim 10^{10}\,\mathrm{M_\odot}$, $z \gtrsim 5$) and at the most massive end ($M \gtrsim 10^{14}\,\mathrm{M_\odot}$, $z < 2$). 
The former part suppresses the low-mass halo abundance by $\sim 10-50\%$ (increasing with redshift), while the latter is almost invisible in Fig.~\ref{fig:fig4} at $M < 10^{14}\, \mathrm{M_\odot}$. 

For the \citet{Planck2014p16} cosmology and a power spectrum with no resolved cut-off, our formula predicts that 
$82.60\%$ of cosmic matter reside in $\left[10^{-6}, 10^{16} \right]\,\mathrm{M_\odot}$ haloes at present, close to the predictions of $77.54\%$ and $77.15\%$ for \citet{ST2002} and \citet{Reed2007}, respectively.\footnote{Note that these fractions depend on the halo definition; for example, for haloes defined by 200 times the cosmic critical density, they will change to 65.45\%, 61.40\%, 61.12\%, respectively. }

\section{Discussion}
\label{sec:discussions}

A physically motivated halo mass function should capture how the non-linear collapse of haloes  departs from the idealized assumptions of the spherical or ellipsoidal collapse models and the excursion set theory. 
Arguably, the empirical corrections introduced in this work are expected when one considers the complications encountered when pushing to extremely low halo masses and high redshifts, while still matching the abundance of massive, late-time haloes on which many previous fits in the literature were originally calibrated. 
Here, we discuss a few possible physical reasons why the classical `universal' form may fail. 

As traditional structure formation theories assume independent random walks in a smoothed density field and a non-evolving collapse threshold, $\delta_\mathrm{c}$ (either constant in the PS model or scale-dependent in the ST model), the well-known `universal' form, $f(\nu)$, has been proposed, where all cosmology and time dependence enters into the formula only via the cosmic variance, $\sigma^2(M)$, and the growth factor, $D(z)$. 
However, this elegant universality may break down in some circumstances:  

\begin{enumerate}

\item {\it Non-Markovian random walks:} for realistic power spectra and real-space filtering, excursion set trajectories are intrinsically non-Markovian, particularly on the small scales relevant to planetary mass haloes. Neglecting these correlations leads to systematic deviations that are often absorbed into empirical fitting functions. When treated self-consistently, however, the non-Markovian nature of the walks restores the validity of the spherical collapse excursion set prediction \citep{Delos2024b}. In practice, most fitting formulae (including the one presented here) should therefore be regarded as effective descriptions within a simplified EPS framework rather than as direct tests of halo collapse mechanisms. 
Nevertheless, the close agreement between the halo mass functions measured in subsampled regions and those from the full PMILL simulation in the overlapping $\ln \nu$ range suggests that $f(\nu)$ inferred from local regions is only weakly affected by the prior excursion set trajectory of the density field, supporting the applicability of the conventional EPS-based framework as an effective approximation in the regime tested here. 

\item {\it Possible environmental impact:} local density, shear and tidal fields might affect the halo mass and abundance. In the regimes tested in this study, however, the EPS-based $\sigma - \nu$ space conversion (i.e., Eq.~1) brings regions with different environmental densities and scales into close agreement in $f(\nu)$ (Fig. \ref{fig:fig1}, \ref{fig:fig2}, and \ref{fig:fig3}), indicating that these side effects  are mostly captured by the EPS theory. 
\end{enumerate}

What our work attempts to do is to calibrate the halo mass function within the conventional framework, where the empirical correction introduced here can be regarded as a redshift-dependent modification of the collapse threshold and mass definition: 

\begin{enumerate}
\item The low-$\nu$ behaviour of the hyperbolic tangent suppression term $\Gamma(\nu, z)$ (further refined with $p_\mathrm{this\ work}$) encodes the fact that the collapse threshold may increase at high redshift and small scales, reducing the abundance of low-mass haloes compared to a universal fit at $z \gtrsim 2$. 

\item The mild change in $G(\nu, z)$ compared to R07 increases the predicted halo mass function at the galaxy cluster-mass end, while calibration through a smaller $c$ and the high-$\nu$ behaviour of $\Gamma(\nu, z)$ further improves the fit across all redshifts. 
As the R07 formula is calibrated using FOF halo masses, this implies that the impact of halo definitions becomes more significant at the high mass end, where the slope of the halo mass function becomes steep. 

\item Since we apply our correction to the R07 functional form, we preserve its success at intermediate masses and low redshifts (where the idealized assumptions above remain adequate), while adding minimal extra flexibility to match the extreme regimes probed by our simulations. 
\end{enumerate}

It is important to stress that the fitting formula presented in this work is calibrated only using simulations of the $\Lambda$CDM cosmology, and therefore may not be the optimal fit for other cosmological models, such as those with a small-scale cutoff in the matter power spectrum associated with different dark matter particles, primordial non-Gaussianity of initial perturbations, different primordial spectral index, or modified background cosmologies (e.g. early dark energy). Nevertheless, previous studies suggest that some of these effects can be incorporated through relatively simple extensions to the standard treatment: 
\begin{enumerate}
\item A cutoff in the primordial power spectrum is well-known to suppress halo formation below the corresponding scale while producing spurious structures that need to be excluded from the analysis \citep[e.g.,][]{Wang2007, Angulo2013}. 
\citet{Bose2016} used the \textsc{coco-cold} and \textsc{coco-warm} simulations \citep{Hellwing2016} and the method of \citet{Lovell2014} to remove spurious haloes. They found that the halo mass function in the cold dark matter case is  predicted well by the ST formula \citep{ST1999}, and that the same holds for the warm dark matter case as long as the window function is switched from a real-space tophat filter to a sharp $k$-space filter. 
Similar results are also reported by  \citet{Benson2013} and \citet{Schneider2013}. 
Since this work derives a formula adapted from the ST formalism, it is likely that the halo mass function in this case can still be well predicted using a sharp $k$-space window function. 
\item Primordial non-Gaussianity tends to have a stronger impact on the number of the most massive haloes at high redshift, with an amplitude that decreases with time, the dimensionless non-linearity parameter, $f_\mathrm{NL}$, and on  smaller scales \citep[e.g.,][]{Matarrese2000, Verde2001, LoVerde2008, Dalal2008, DAmico2011}. 
For example, using N-body simulations, \citet{Fiorino2025} showed that, with $f_\mathrm{NL}=100$, the halo mass function at $z=1$ increases by $\sim28\%$ and $4\%$ at $M=10^{15}$, $10^{14}\,\mathrm{M}_\odot$, respectively, and these numbers fall to $4\%$ and $<1\%$ at $z=0$. 
\citet{Desjacques2009} suggested that the correction to the halo abundance at different times and mass scales can be well fitted by a function of $\nu$; 
\citet{Fiorino2025} provided another fitting formula to include the impact of different halo definitions. 
Given that recent Planck results  \citep{PlanckCollaboration2020} have constrained local, equilateral, and orthogonal non-Gaussianity to $f_\mathrm{NL}^\mathrm{loc}=-0.9 \pm 5.1$, $f_\mathrm{NL}^\mathrm{equil}=-26 \pm 47$, $f_\mathrm{NL}^\mathrm{ortho}=-38 \pm 24$, respectively, we  expect that this effect would be even smaller, and refer interested readers to the formulae listed above to correct the halo mass function. 
\end{enumerate}

\section{Conclusions}
\label{sec:conclusions}

In this work, we have presented the full halo mass function in the $\Lambda$CDM cosmology over an unprecedented dynamic range, based on the VVV and P-Millennium simulations, as well as the new simulation VVV2.8 (see Section~\ref{sec:simulation}).
By introducing a subsampling technique that reconstructs a global halo mass function from biased underdense zoomed regions, we measured the halo mass function robustly for haloes from planetary to galaxy cluster mass, and from $z=30$ to the present. 

From  Figures~\ref{fig:fig1}, \ref{fig:fig2} and \ref{fig:fig3} we find that the simple \citet{PS1974} (PS) mass function is relatively inaccurate whereas the more complicated \citet{Reed2007} (R07) formula greatly improves the agreement with the simulation results at low redshift; however,  it deviates by as much as a factor of $\sim 2$ when extrapolated to high redshift and low mass. We introduce  corrections to the R07 formula (see Section~\ref{Empirical}): a suppression term, $\Gamma(\nu, z)$, whose low-$\nu$ behaviour lowers the predicted number of low mass haloes at high redshift (especially for $M \lesssim 10^{10}\,\mathrm{M}_\odot$ and at $z \gtrsim 2$); and a mild calibration of the Gaussian correction term, $G(\nu, z)$, together with a smaller value of the parameter $c$ and the high-$\nu$ behaviour of $\Gamma(\nu, z)$, which improve the fit for the most massive haloes. 
These corrections preserve the success of the R07 formula within its tested ranges, while achieving $\sim 2-7\%$ accuracy over the entire range of mass and redshift  ($10^{-6}-10^{15.5}\,\mathrm{M}_\odot$, $z=0-30$). 

The new fitting formula provides the evolution of halo abundance in the $\Lambda$CDM model across an unprecedented dynamic range (Fig.~\ref{fig:fig4}). From the limited tests we have performed (see Appendix~\ref{App-B}), it is also accurate for models with modest deviations from $\Lambda$CDM in the values of some of the cosmological parameters. Thus, our formula 
is a practical and accurate tool for structure formation studies covering the full history, from the first mini-haloes to present-day clusters, and is useful for a broad range of applications such as dark matter annihilation, cluster cosmology and high-$z$ galaxy modelling.

\section*{Acknowledgements}
We thank the anonymous referee for a very constructive report which helped to improve our manuscript. 
We thank the FLARES team for providing access to their data, and Xiaoyue Cao, Willem Elbers, and Kai Zhu for useful discussions and comments. 
We acknowledge the support from the National Natural Science Foundation of China
(No. 12473007, 12588202, 12473015), Beijing Natural Science Foundation (QY23018), and the China Manned Space Program with grant No. CMS-CSST-2025-A03 and No. CMS-CSST-2025-A09. 
SB is supported by the UK
Research and Innovation (UKRI) Future Leaders Fellowship [grant numbers
MR/V023381/1 and UKRI2044]. CSF acknowledges support by the European Research
Council (ERC) through Advanced Investigator grant, DMIDAS (GA
786910). ARJ is supported by the STFC consolidated grant ST/X001075/1.
JW acknowledges the support of the
research grants from the Ministry of Science and Technology
of the People’s Republic of China (No. 2022YFA1602901), 
the China Manned Space Project (No. CMS-CSST-2021-B02), 
and the CAS Project for Young Scientists in Basic Research (Grant No. YSBR-062). 
This work used the DiRAC@Durham facility managed by the
Institute for Computational Cosmology on behalf of the STFC DiRAC HPC
Facility (www.dirac.ac.uk). The equipment was funded by BEIS capital
funding via STFC capital grants ST/K00042X/1, ST/P002293/1,
ST/R002371/1 and ST/S002502/1, Durham University and STFC operations
grant ST/R000832/1. DiRAC is part of the UK National e-Infrastructure.

\section*{Data availability}
The data used in this paper will be shared upon reasonable request
to the corresponding author. 

\bibliographystyle{mnras}
\bibliography{main} %

\appendix

\section{Halo mass function for alternative halo definitions}
\label{App-A}

In the main text, we define the halo mass using the radius within which the average matter density equals 200 times the cosmic mean matter density. 
Meanwhile, various alternative halo definitions exist in the literature, such as using 200 times the critical density (such a halo mass is denoted as $M_\mathrm{200c}$ hereafter), or adopting a spherical-collapse parameter $\Delta_\nu$ times the critical density of the universe \citep{Eke1996, Bryan&Norman1998}. 

At low redshifts ($z \lesssim 2$), the critical density of the universe is much higher than the mean density, which results in smaller halo masses when using $M_\mathrm{200c}$ instead of $M_\mathrm{200}$, and it correspondingly shifts the halo mass function toward the low mass end. 
Here we test a method to map between halo mass functions defined under different halo-mass conventions, using halo density profiles and mass -- concentration relations. 

\citet{Wang2020} found that haloes with masses across such a wide dynamic range can be well described by the NFW or Einasto profiles \citep{Navarro1996, Einasto1965}: 
\begin{equation}
    \rho_\mathrm{NFW}(r) = \rho_\mathrm{s} \, r_\mathrm{s}^3 \, r^{-1} \, (r + r_\mathrm{s})^{-2}, 
\end{equation}
\vspace{-0.5cm}
\begin{equation}
    \rho_\mathrm{Einasto}(r) = \rho_\mathrm{-2} \exp \left[-2\alpha^{-1}((r / r_{-2})^\alpha - 1) \right], 
\end{equation}
where $\rho_\mathrm{s}$ and $r_\mathrm{s}$ are the characteristic density and scale radius, $\rho_{-2}$ and $r_\mathrm{-2}$ are the density and radius at which the logarithmic slope is $-2$, and $\alpha$ is a shape parameter. 
For the mass -- concentration relation, we use the analytic model of \citet[][]{Ludlow2016} (i.e., their Eqs. 6-7) for the applicability at different redshifts. 
Following \citet[][]{Ludlow2016}, we set $\alpha=0.18$. 

\begin{figure*}
    \centering
    \includegraphics[width=2.0\columnwidth]{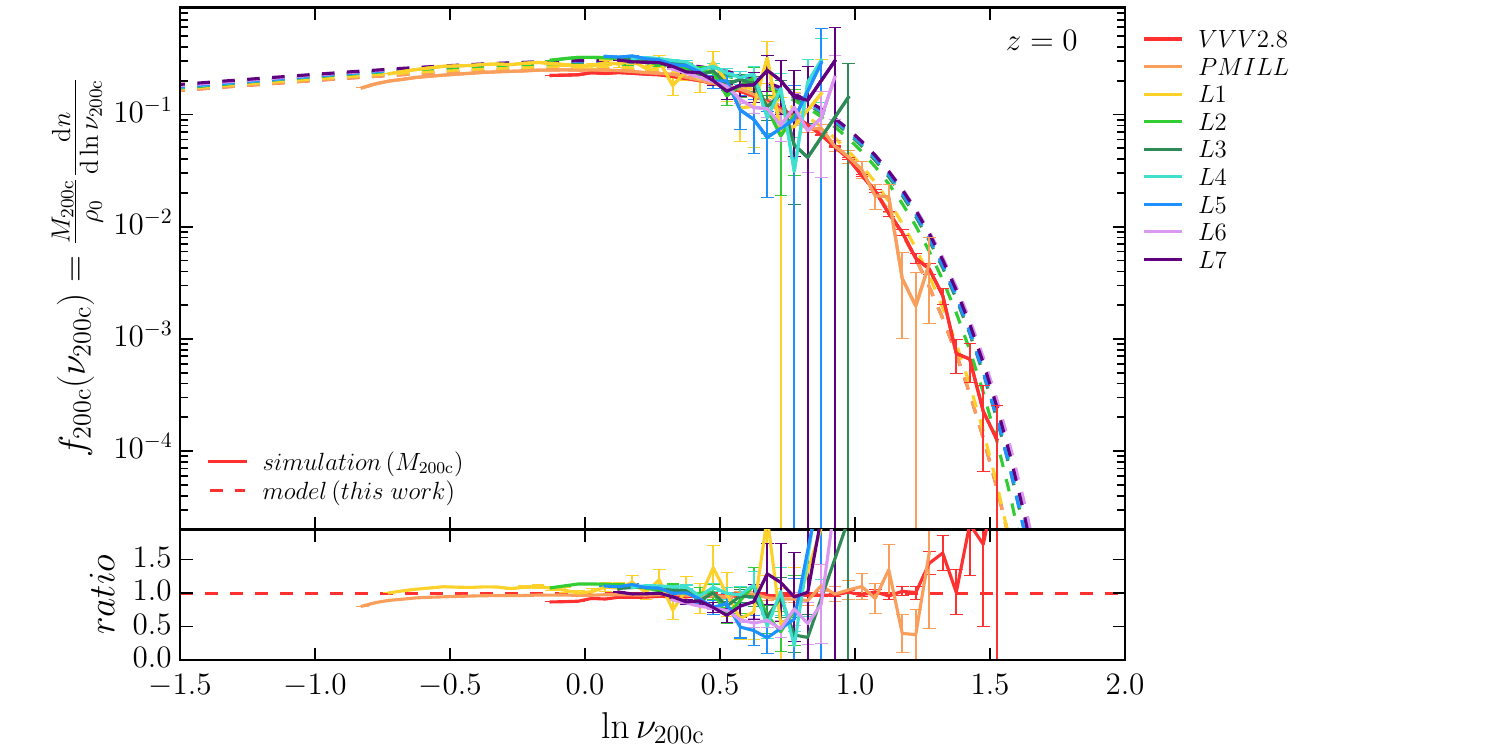}
    \vspace{-0.10cm}
    \caption{Similar to Fig.~\ref{fig:fig1}, but use $M_\mathrm{200c}$ instead. Our predictions are extended to this mass definition using the Einasto profile \citep{Einasto1965} and the mass -- concentration relation as modelled in \citet{Ludlow2016}, and therefore depend on the halo mass probed at different simulation levels. We only plot the halo mass function from original simulation volumes, and the predictions from our fit for simplicity. The good agreement between the halo mass function of the simulation and our model demonstrates that our formula can be accurately extended to other halo definitions. }
    \label{fig:figA1}
\end{figure*}

Using the assumptions above, we build a numerical relation of $M_{200} - M_\mathrm{200c}$, subsequently $\sigma(M_{200}) - \sigma_\mathrm{200c}(M_\mathrm{200c})$ and $\nu(M_{200}) - \nu_\mathrm{200c}(M_\mathrm{200c})$. The model prediction thereby can be transformed to: 
\begin{equation}
\label{eq:A4}
    f_\mathrm{200c}(\nu_\mathrm{200c}) = \frac{M_\mathrm{200c}}{M_\mathrm{200m}} \left(\frac{\mathrm{d}\ln \nu_\mathrm{200c}}{\mathrm{d}\ln \nu}\right)^{-1}f(\nu). 
\end{equation}
It is worth noting that this formula now depends on the physical halo mass scale, as introduced by the mass – concentration relation.
In Fig. \ref{fig:figA1}, we find very good agreement between the simulation and our model predictions, confirming that our formula is applicable across different halo definitions. 

We note that the halo mass function may also depend on the choice of halo finder. Previous studies comparing halo finders \citep[e.g.,][]{Knebe2011} generally find good agreement for the abundance of central haloes, while larger differences arise primarily for subhaloes and poorly resolved objects. 
More recently, \citet{ForouharMoreno2025} found that the abundances of central haloes identified with \textsc{subfind} and \textsc{hbt-herons} (using the $M_\mathrm{200c}$ definition) agree to within $\sim 1\%$. Since this paper only consider central haloes with at least 50 dark matter particles and adopt a spherical overdensity mass definition, we expect the sensitivity of our results to the choice of halo finder to be relatively small.

\section{Halo mass function for different cosmological parameters}\label{App-B}

\begin{figure*}
    \centering
    \includegraphics[width=2.0\columnwidth]{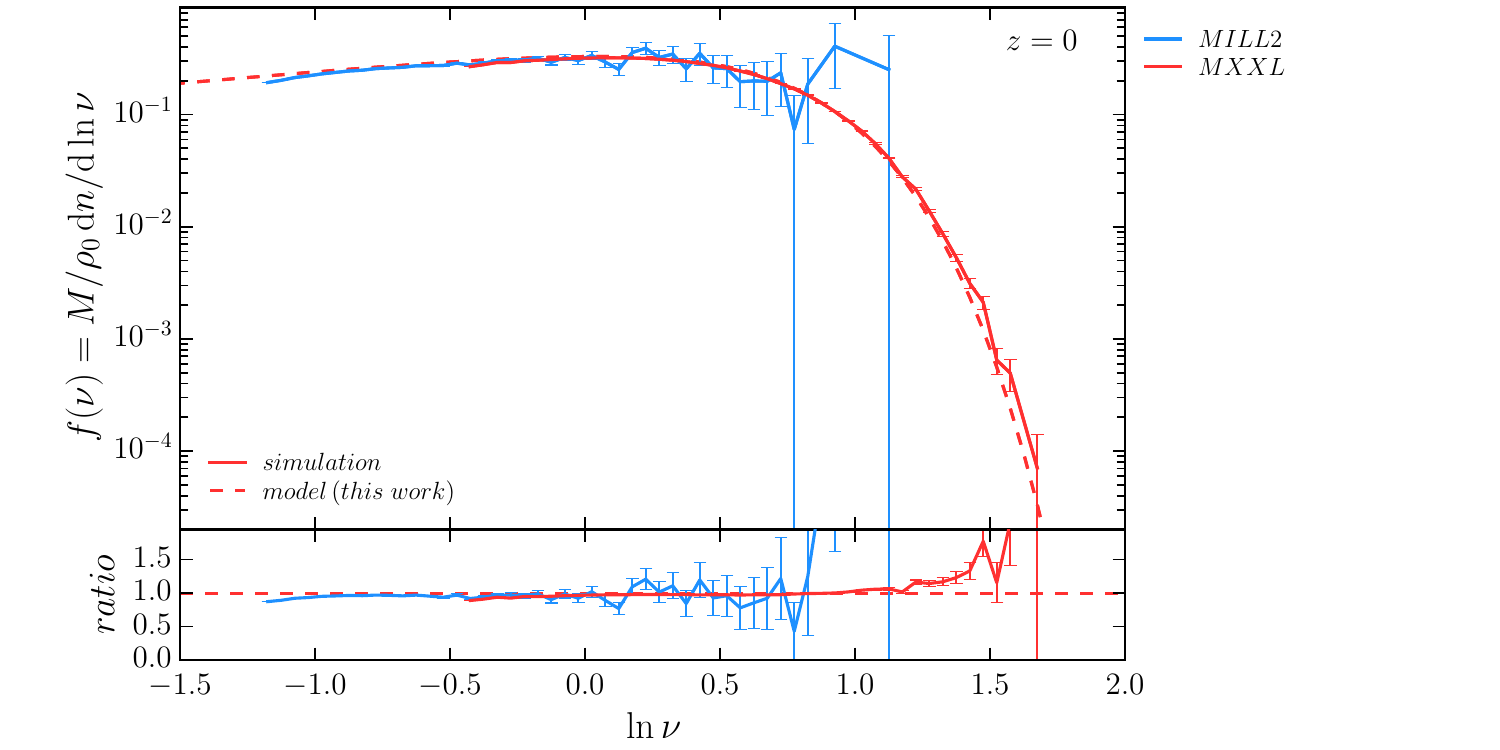}
    \vspace{-0.10cm}
    \caption{Similar to Fig.~\ref{fig:fig1}, but using the Millennium-2 \citep[MILL2, blue solid line]{Boylan-Kolchin2009} and Millennium-XXL \citep[MXXL, red solid line]{Angulo2012} simulations with the WMAP-1 cosmology \citep{Spergel2003} instead. The good agreement suggests that our formula (red dashed line) can be accurately applied in other cosmologies. }
    \label{fig:figB1}
\end{figure*}

In order to test how well our formula can be applied to different cosmologies, we examine in Fig. \ref{fig:figB1} the Millennium-2 \citep[MILL2]{Boylan-Kolchin2009} and Millennium-XXL \citep[MXXL]{Angulo2012} simulations, which adopt the WMAP-1 cosmology \citep[$\Omega_\mathrm{m}=0.25$, $\Omega_\mathrm{\Lambda}=0.75$, $h=0.73$, $\sigma_8=0.9$]{Spergel2003} and differs somewhat from the \citet{Planck2014p16} cosmology used in our main text. 
Our model prediction is well aligned with these simulations, proving its applicability across different cosmologies. 

\section{Tests of the subsampling method}
\label{App-C}

Based on EPS theory, we propose a novel subsampling method to probe the halo mass function to extremely low masses, building on the premise that the random walk from the extracted subregions to halo collapse is independent of the random walk from the VVV regions to those subregions. 
Here we use the dark-matter-only version of the FLARES simulations \citep{Lovell2021} to test whether and to what extent the underlying assumptions of this method hold. 
The FLARES simulations consist of 40 zoom-in runs selected from a $3.2\,\mathrm{cGpc}$ box, and each run represents a spherical region with a radius of $20.66\,\mathrm{cMpc}$, resolved with a dark matter particle mass of $\sim10^7\,\mathrm{M_\odot}$. 
Their rms overdensities range from $-5.29$ to $5.62$, sampling a wide variety of environments and thus making them ideal for testing the applicability of our method. 

\begin{figure*}
    \centering
    \includegraphics[width=2.0\columnwidth]{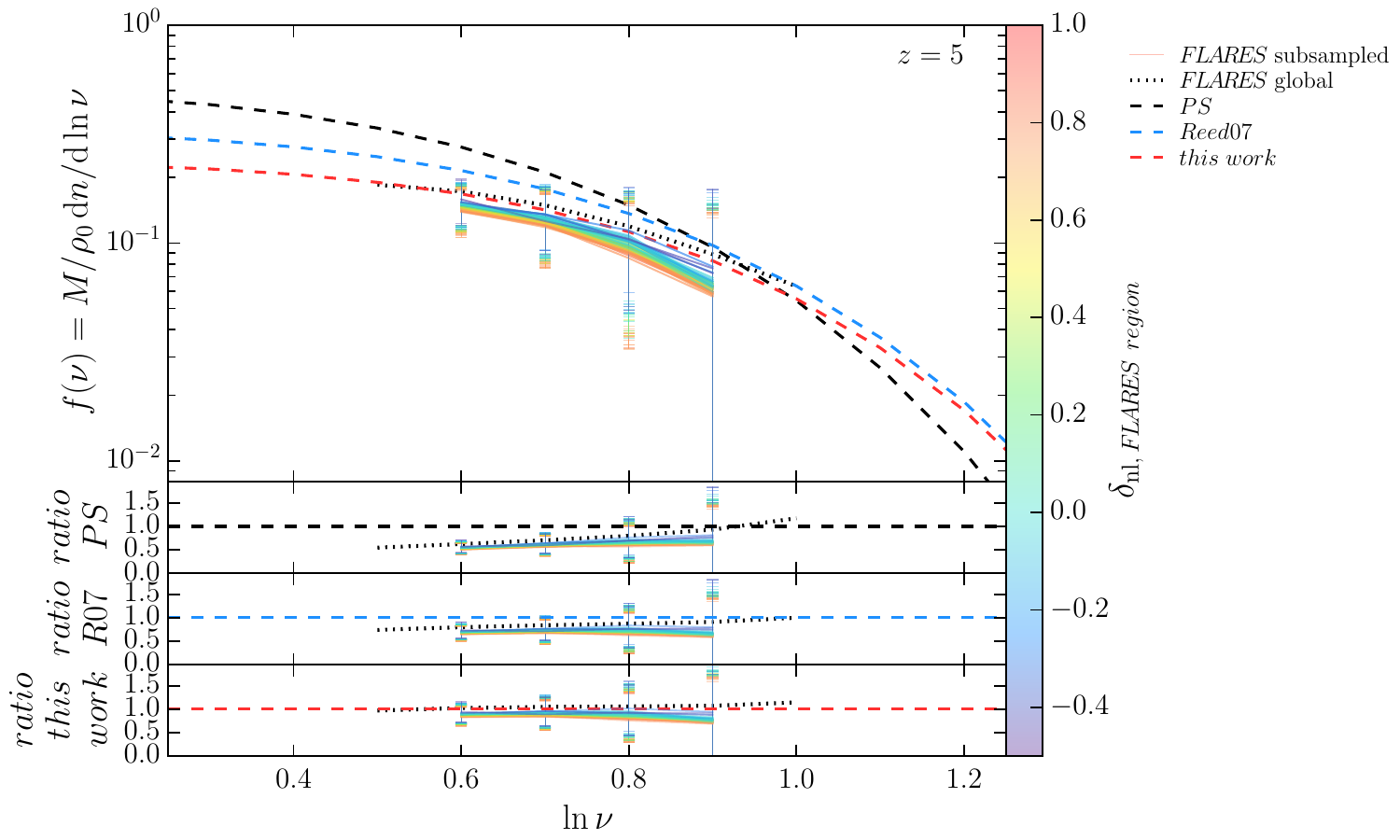}
    \caption{Similar to Fig.~\ref{fig:fig1}, but extracting subregions with cosmic mean density from the 40 zoom-in runs in the FLARES simulations \citep[dark-matter-only, ][]{Lovell2021}. Each solid line shows the halo mass function of the subregions extracted from one FLARES run, with the colour indicating the overdensity of the corresponding FLARES region. The dotted line shows the global halo mass function reproduced by combining all 40 regions using the weighting method of \citet{Lovell2021}. The reasonable agreement among the results from different environments, the global halo mass function, and our fit suggests our subsampling method can reproduce the global halo mass function from an extremely underdense region (e.g., those in the VVV simulations) with relatively little bias. }
    \label{fig:figC1}
\end{figure*}

We use the snapshots at $z=5$ (approximately the endpoint of the simulations) and implement the same subsampling method, extracting spherical subregions with overdensity $\delta=0$ and a radius of $2.23\,\mathrm{cMpc}$ from each run. We then show the median and scatter of the halo mass function among the subregions in Fig. \ref{fig:figC1}. The result for each run is plotted individually, with the colour representing the overdensity of the corresponding FLARES region. We also use the weighting method of \citet{Lovell2021} to combine all 40 regions and reconstruct a global halo mass function shown by the dotted line. 

The results from the subsampling in different FLARES regions, the global weighted reconstruction, and our fit are broadly consistent with each other, suggesting that our method can largely separate out the bias associated with the underdense VVV environment and thereby infer the global halo mass function with reasonable accuracy. 
Note that we do find a mild trend, especially at the high $\nu$ end, where the subsampled halo mass function slightly decreases with the overdensity of the FLARES regions. 
This might be related to the window function: strictly speaking, only a sharp-$k$ filter can guarantee completely independent increments of the random walk, while the tophat filter used in this work can introduce correlations between different scales and therefore retain some dependence on the preceding walk from the FLARES regions to the subregions. 
Taking an overdense FLARES region as an example, the random walk needs to proceed `downhill' to reach subregions with cosmic mean density, which could be weakly correlated with the subsequent excursion toward halo collapse. 
This may also explain why the low $\nu$ end show a weaker correlation with the densities of regions. 
These residual trends, however, do not significantly affect our fit, which remains in good agreement with the reconstructed global halo mass function in Fig. \ref{fig:figC1}.

\bsp	
\label{lastpage}
\end{document}